\documentclass[10pt,journal,compsoc]{IEEEtran}

\usepackage{cite}
\usepackage{amsmath,amssymb,amsfonts}
\usepackage{algorithmic}
\usepackage{graphicx}
\usepackage{textcomp}
\usepackage{xcolor}
\usepackage{subfigure}
\usepackage{float}
\usepackage{bm}
\usepackage{soul}
\usepackage{comment}

\def\BibTeX{{\rm B\kern-.05em{\sc i\kern-.025em b}\kern-.08em
    T\kern-.1667em\lower.7ex\hbox{E}\kern-.125emX}}

\ifCLASSINFOpdf
\else
\fi

\begin{document}

\newcommand{\systemname}{HAR-SAnet$ $}

\title{RF-Based Human Activity Recognition Using Signal  Adapted Convolutional Neural Network}

\author{Zhe Chen\IEEEauthorrefmark{1},
		Chao Cai\IEEEauthorrefmark{1}, %
		Tianyue Zheng\IEEEauthorrefmark{1}, 
		Jun Luo\IEEEauthorrefmark{1},
		Jie Xiong\IEEEauthorrefmark{3}, Xin Wang \IEEEauthorrefmark{4}
\IEEEcompsocitemizethanks{\IEEEcompsocthanksitem Zhe Chen, Chao Cai, Tianyue Zheng, Jun Luo are with School of Computer Science and Engineering, Nanyang Technological University, Singapore, Singapore.\protect\\
E-mail: \{chen.zhe, chris.cai, tianyue002, junluo\}@ntu.edu.sg
\IEEEcompsocthanksitem Jie Xiong is with University of Massachusetts Amherst, Amherst, USA. E-mail:jxiong@cs.umass.edu
\IEEEcompsocthanksitem Xin Wang is with School of Computer Science, Fudan University, Shanghai 200433, China, and also with Shanghai Key Laboratory of Intelligent Information Processing, Shanghai 200433, China. E-mail:xinw@fudan.edu.cn}
\thanks{This work was supported in part by AcRF Tier 1 Grant RG17/19 and the National Natural Science
Foundation of China (Project No. 61971145). }
}

\IEEEtitleabstractindextext{%
\begin{abstract}
Human Activity Recognition (HAR) plays a critical role in a wide range of real-world applications, and it is traditionally achieved via wearable sensing. Recently, to avoid the burden and discomfort caused by wearable devices, device-free approaches exploiting Radio-Frequency (RF) signals arise as a promising alternative for HAR. Most of the latest device-free approaches require training a large deep neural network model in either time or frequency domain, entailing extensive storage to contain the model and intensive computations to infer human activities. Consequently, even with some major advances on device-free HAR, current device-free approaches are still far from practical in real-world scenarios where the computation and storage resources possessed by, for example, edge devices, are limited.
To overcome these weaknesses, we introduce HAR-SAnet which is a novel RF-based HAR framework. It adopts an original signal adapted convolutional neural network architecture: instead of feeding the handcraft features of RF signals into a classifier, HAR-SAnet fuses them adaptively from both time and frequency domains to design an end-to-end neural network model. We apply point-wise grouped convolution and depth-wise separable convolutions to confine the model scale and to speed up the inference execution time. The experiment results show that the recognition accuracy of HAR-SAnet substantially outperforms the state-of-the-art algorithms and systems.
\end{abstract}

}

\maketitle

\IEEEdisplaynontitleabstractindextext

\IEEEpeerreviewmaketitle

\IEEEraisesectionheading{\section{Introduction}\label{sec:introduction}}

\IEEEPARstart{H}{uman} Activity Recognition (HAR) has attracted a significant amount of attentions in the past decade due to its great value in a wide range of real-world applications, such as health care \cite{b24, b48}, fall detection \cite{b20,b25,b28}, and smart home \cite{b1,b49}. 
There are generally two types of solutions for HAR: \textit{device-based} and \textit{device-free}. Device-based solutions rely mostly on wearable devices such as smartphones and smart watches. However, these solutions often cause discomfort and extra burden. 
To overcome the weaknesses, device-free solutions utilizing cameras and Radio-Frequency (RF) signals have later come into view. Recently, camera-based HAR systems have achieved successes in several outdoor scenarios thanks to deep learning, but they may not be well-accepted in indoor environments due to the severe privacy concerns~\cite{b1}. Different from camera-based solutions, 
RF-based approaches do not raise privacy concerns, and are not affected by temperature or lighting conditions. Therefore, RF-based solution has become a promising candidate for indoor HAR, leading to a large amount of research contributions recently~\cite{b2,b3,b4,b5,b6,b8,b9,b10,b11}. %

The basic principle of RF-based HAR systems is that the propagation paths of RF signals are affected by human body movement, causing the reflected signals to exhibit distinct features pertaining to different activities. Consequently, we can exploit these unique features to distinguish different activities, hence significant progress on RF-based HAR has been achieved in the past few years~\cite{b2,b3,b4,b5,b6,b7,b8,b9,b11,b13,b14}. 
Among all wireless signals used for HAR, Wi-Fi is the most popular one owning to the ubiquitous deployment~\cite{b2,b3,b4,b5,b6,b8,b9,b10}. %

Though promising, several major challenges still exist with the state-of-the-art Wi-Fi-based approaches, hindering the adoption of these systems in real life: 
\begin{itemize}
	\item Narrow Wi-Fi channel bandwidth leads to limited resolution in differentiating  activity patterns.
	\item While preprocessing the raw signal collected from the hardware helps removing the signal noise, the important signal feature containing the activity information may also get lost.
	\item Low computation capability edge devices have difficulty to achieve real-time HAR.
	\item Edge devices with limited memory cannot support a large neural network running on it.
\end{itemize}
Furthermore, although Wi-Fi infrastructure is ubiquitously deployed, the CSI information employed for HAR cannot be retrieved from most commodity Wi-Fi hardware but only from the Intel 5300 and several specific Atheros Wi-Fi cards, limiting the practical adoption of Wi-Fi-based approaches.

In this paper, to address the above challenges, we employ a Commercial Off-The-Shelf (COTS) Ultra-Wide Band~(UWB) radio module for HAR. Compared with Wi-Fi, UWB radio has a much larger channel bandwidth and thus a much higher time resolution. We show that the UWB module has a comparable cheap price as the Wi-Fi card but can achieve a much better performance in terms of both HAR accuracy and robustness.  %

For HAR, another big issue is how to extract stable and unique features related to each activity. However, these features depend highly on the individuals: body size and personal habits can cause large variations~\cite{b13,b14} in the features extracted. Fortunately, resorting to Convolutional Neural Network (CNN), the complex features of various types of signals such as images and video have been effectively extracted~\cite{b15,b16}. In other words, CNN opens a new paradigm for HAR, whose power has been demonstrated in various RF-based HAR systems~\cite{b8,b9,b11,b12,b16,enhancing}. Whereas most of exiting systems consider either time or frequency domain information for HAR, we propose to employ both time and frequency domain information to achieve more accurate and more robust performance.
Essentially, we employ two CNN branches to learn the feature representations of time and frequency domains, respectively. Then, RF features from both time and frequency domains are fused together to infer human activities.
Therefore, our system utilizes the information extracted from RF signals to the fullest extent. %

A big issue hindering the real-life adoption of CNN-based approaches is the high computational cost and large storage memory requirement. %
In the Internet-of-Things (IoT) era, the resource-constrained edge nodes or devices usually do not have such a powerful computational power and the storage memory is also limited. Take the popular Raspberry~Pi~Zero~W as an example, it has a 1~\!GHz, single-core CPU (ARMv6) and 512~\!MB RAM~\cite{b50}. Therefore, implementing the proposed RF-based CNN model on a resource-constrained edge device poses a significant challenge. To this end, we customize each CNN block in our signal adapted CNN model structure. In contrast to conventional camera images that all unoccluded key-points of the human are recorded, RF signals only get reflected from a subset of the human body parts and the number of reflection points is usually less than seven~\cite{b55}. Due to the sparsity of RF signals, we employ dilated convolutions~\cite{b43} to encode more effective features from RF spectrograms. Moreover, we avoid large CNN model block, such as ResNet~\cite{b15} or full connections~\cite{b47} that incur larger computation and storage overhead. Instead, we resort to efficient designs such as channel split, grouped convolutions, depth-wise  convolutions, and point-wise convolutions. As a result, \systemname\ contains only lightweight components to efficient reduce both computation and storage complexity, making our design work well on the less powerful edge devices.

We design \systemname\ and evaluate its performance on ARM-based edge devices. We productize our system and it is now ready for sale~\cite{wirush}. We test \systemname\ with over thirty persons aged 20-45 years performing seven types of activities, including bending, falling, lying down, standing up, sitting down, squatting down, and walking. The experiment results show that \systemname\ not only demonstrates high recall and precision, but also achieves HAR in real time %
with a small millisecond level end-to-end latency on ARM-based resource-constrained edge devices. To summarize, we make the following contributions.
\begin{itemize}
	\item To the best of our knowledge, we propose the first real-time HAR prototype involving a carefully designed hardware and a signal processing pipeline tailored to resource-constrained edge devices. 
	\item To improve the accuracy, our signal adapted neural network model innovates in taking into account information from both time and frequency domains. 
	\item We design, implement, and productize \systemname. Extensive experiments are conducted to evaluate the system performance in diverse environments.
	The results show that our system can achieve high accuracy for HAR in real-world environments. %
\end{itemize}

The paper is organized as follows. In Sec.~\ref{sec:challenges}, we explain the practical challenges of existing RF-based HAR systems. Then, we describe the details of system design in Sec.~\ref{sec:design}. In Sec.~\ref{sec:impl_eval}, we present the implementation details as well as the experiment results. 
The related work is discussed in Sec.~\ref{sec:rw}, followed by a conclusion in Sec.~\ref{sec:conclusion}.

\section{Wi-Fi or UWB?} \label{sec:challenges}
In this section, we show the practical challenges with WiFi-based HAR systems, and we also briefly demonstrate the superiority of adopting UWB-based technologies. 

\subsection{Limited Resolution}
802.11 Wi-Fi is a narrowband technology employing only 20MHz-80MHz channel for data communication. To differentiate human activities, time-frequency analysis such as Short-Time Fourier Transform (STFT), and Wavelet Transform (WT) are used to produce a time-frequency spectrogram to differentiate different human activities.   %
However, the channel bandwidth fundamentally limits the time domain signal resolution: with a larger bandwidth, the signals have a higher chance to be separated in time domain~\cite{b56}. Consequently, signals reflected from different body parts have a higher chance to be separated and richer information about each body parts can be obtained.
Nevertheless, even with the latest IEEE 802.11ac Wi-Fi standard~\cite{b57}, the channel bandwidth is still quite limited~(80MHz) given the need for fine-grained HAR.      %

\begin{figure}[h]
	\centering
	\subfigure[Sitting down.]
	{
		\centering
		\includegraphics[width=0.46\linewidth]{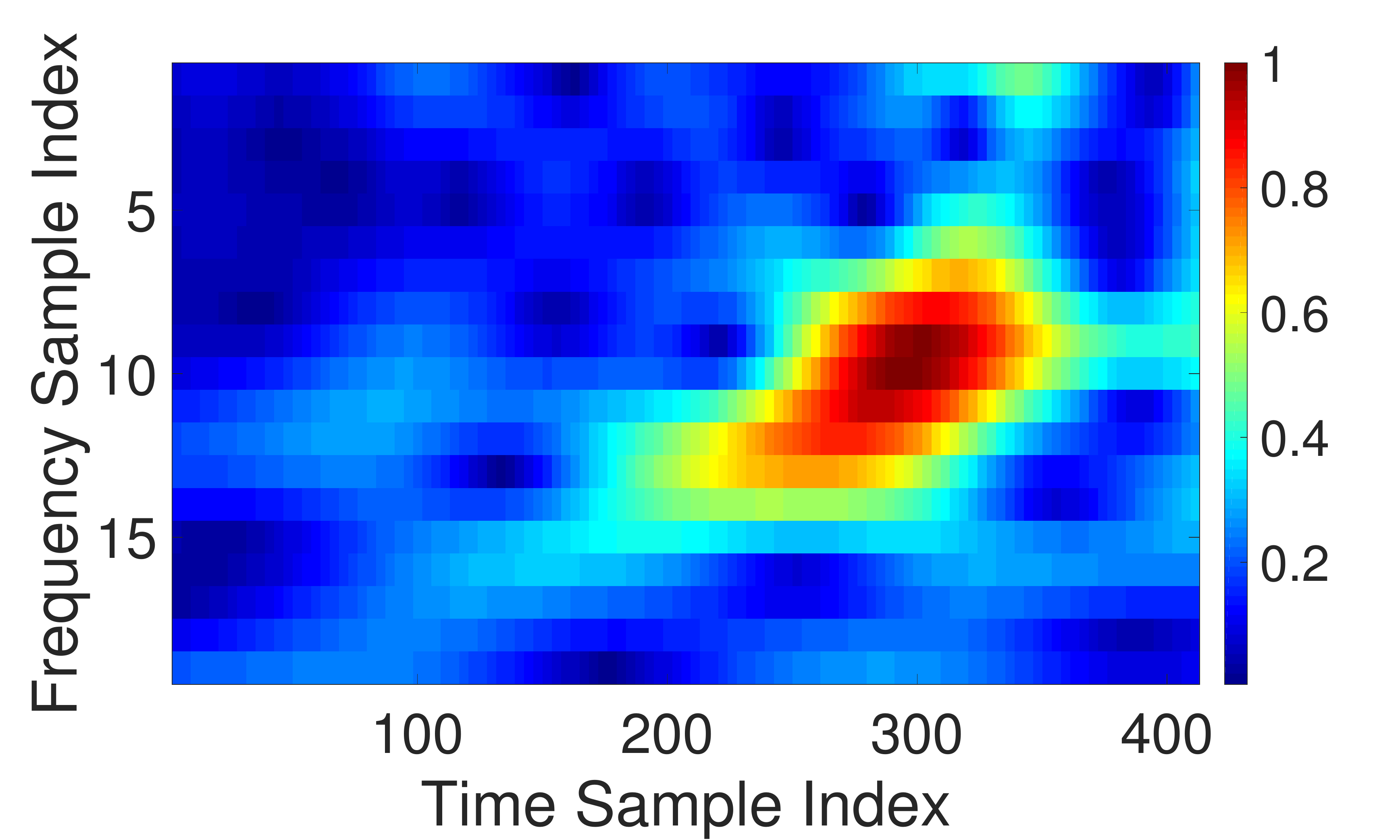}
		\label{fig:sit_tf}	
	}
	\subfigure[Squatting down.]
	{
		\centering
		\includegraphics[width=0.46\linewidth]{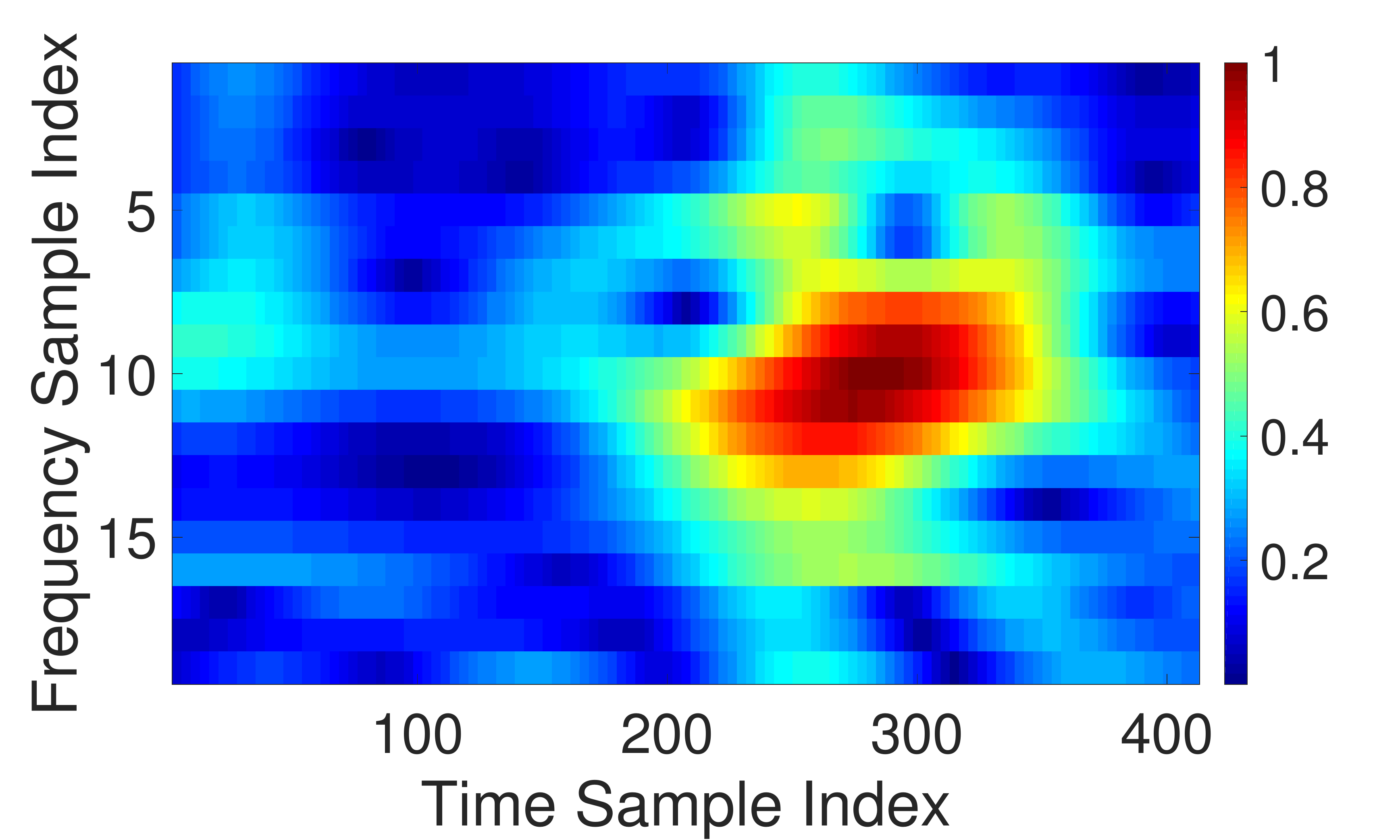}
		\label{fig:squat_tf}	
	}
 	\caption{The time-frequency spectrograms of two similar activities obtained via WT based on Wi-Fi CSI.}
	\label{fig:wifi_act}
\end{figure}

To better understand the limitation of Wi-Fi in terms of sensing resolution, we use the  Intel 5300 Wi-Fi card~\cite{b31} to collect a few data samples of activities at a 400~\!Hz sampling rate. We employ WT method to transform the signal to 
time-frequency spectrograms. %
Fig.~\ref{fig:sit_tf} and Fig.~\ref{fig:squat_tf} illustrate such spectrograms of two activities: sitting down and squatting down. We can see that both spectrograms of these two activities have very similar ``hot'' zones with higher energy. As those hot zones are  to be extracted as features via deep learning, it is error-prone to distinguish these two activities using these spectrograms as input, especially when there is interference and noise. Therefore, Wi-Fi, with a narrowband, can hardly separate motions from different human body parts. Since the narrow bandwidth leads to limited time resolution, Wi-Fi-based systems usually process the signal input only in frequency domain.

\begin{figure}[t]
	\centering
	\subfigure[Sit down: the 22-nd fast-time index.]
	{
		\centering
		\includegraphics[width=0.46\linewidth]{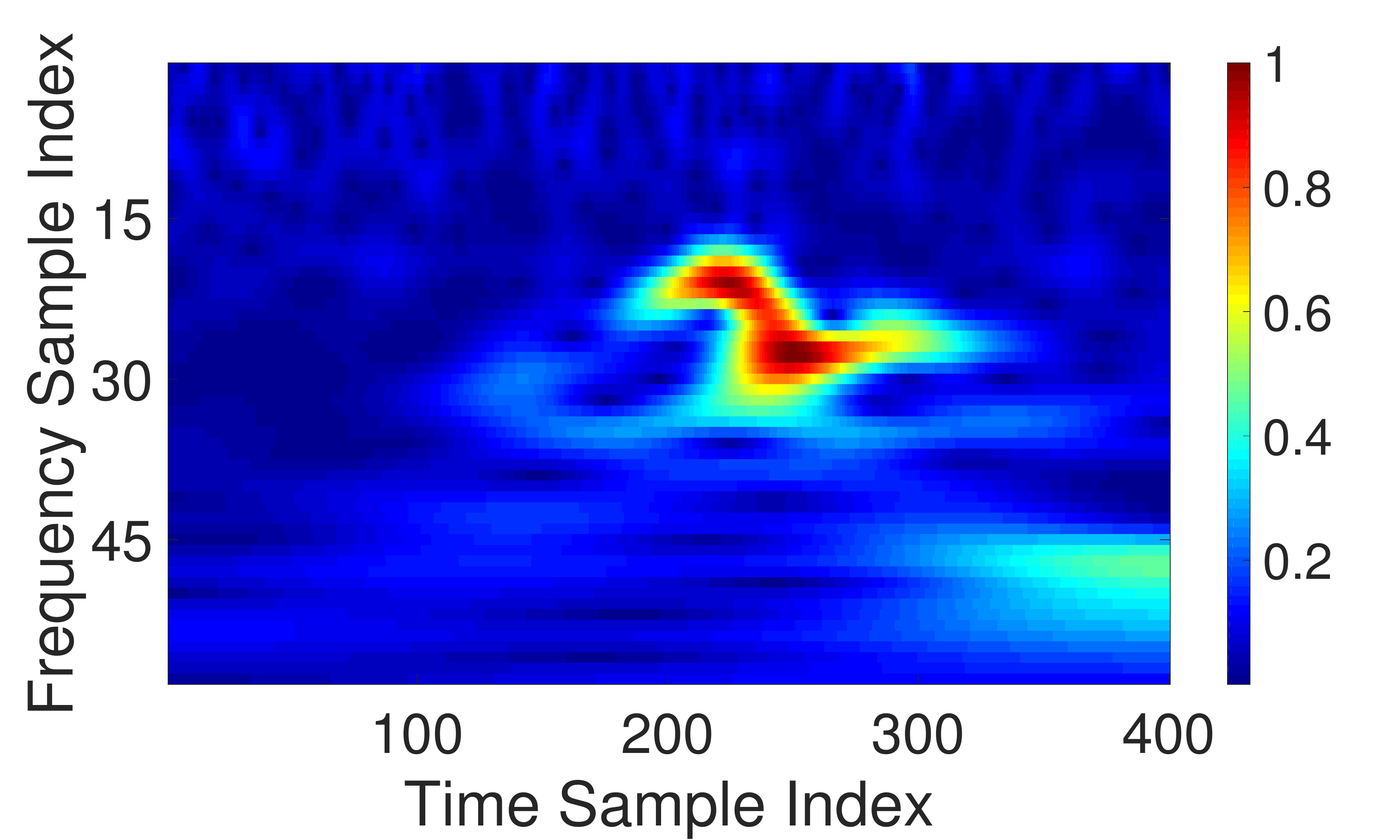}
		\label{fig:sit_uwb_cwt_1}	
	}
	\subfigure[Squat down: the 22-nd fast-time index.]
	{
		\centering
		\includegraphics[width=0.46\linewidth]{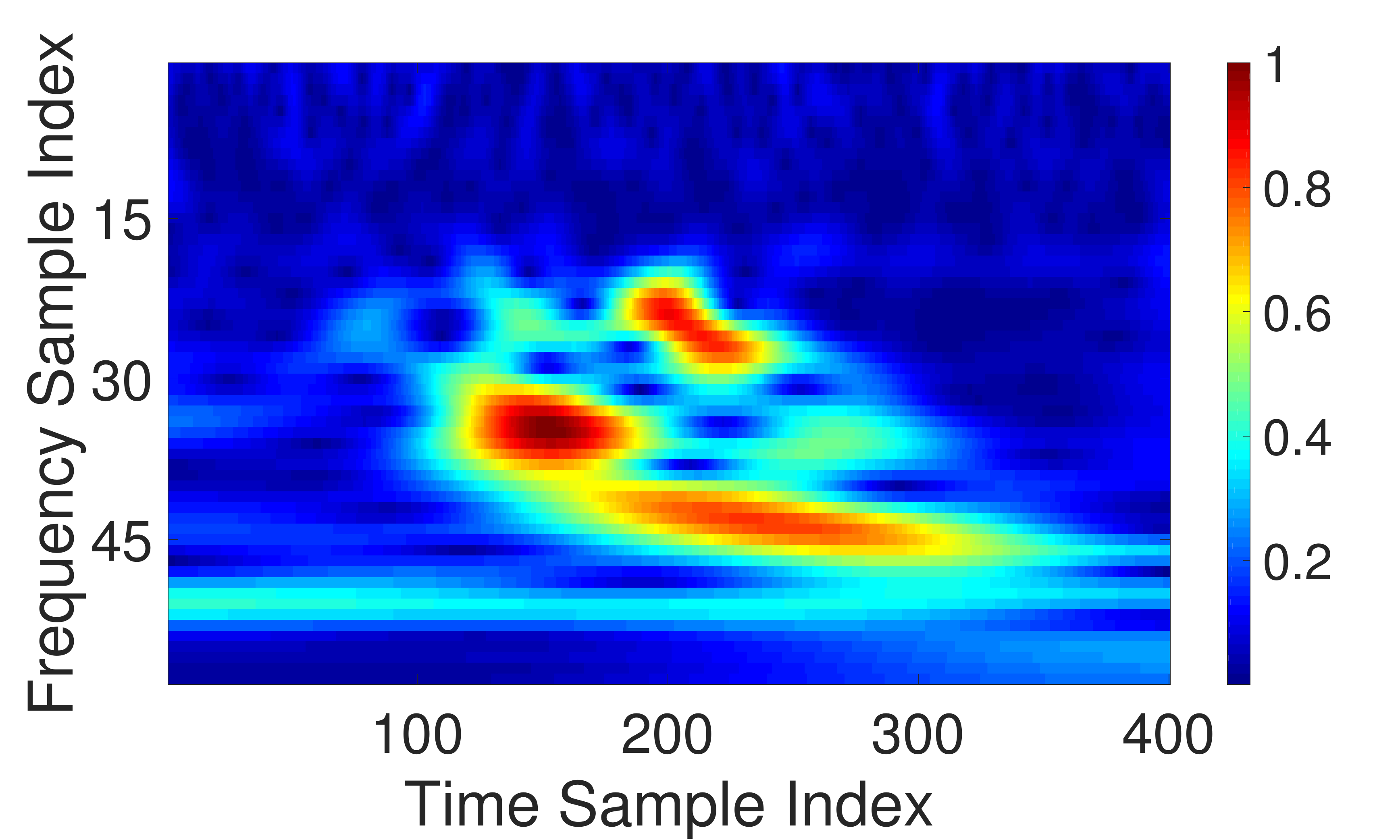}
		\label{fig:sqaut_uwb_cwt_1}	
	}
	\subfigure[Sit down: the 35-th fast-time index.]
{
	\centering
	\includegraphics[width=0.46\linewidth]{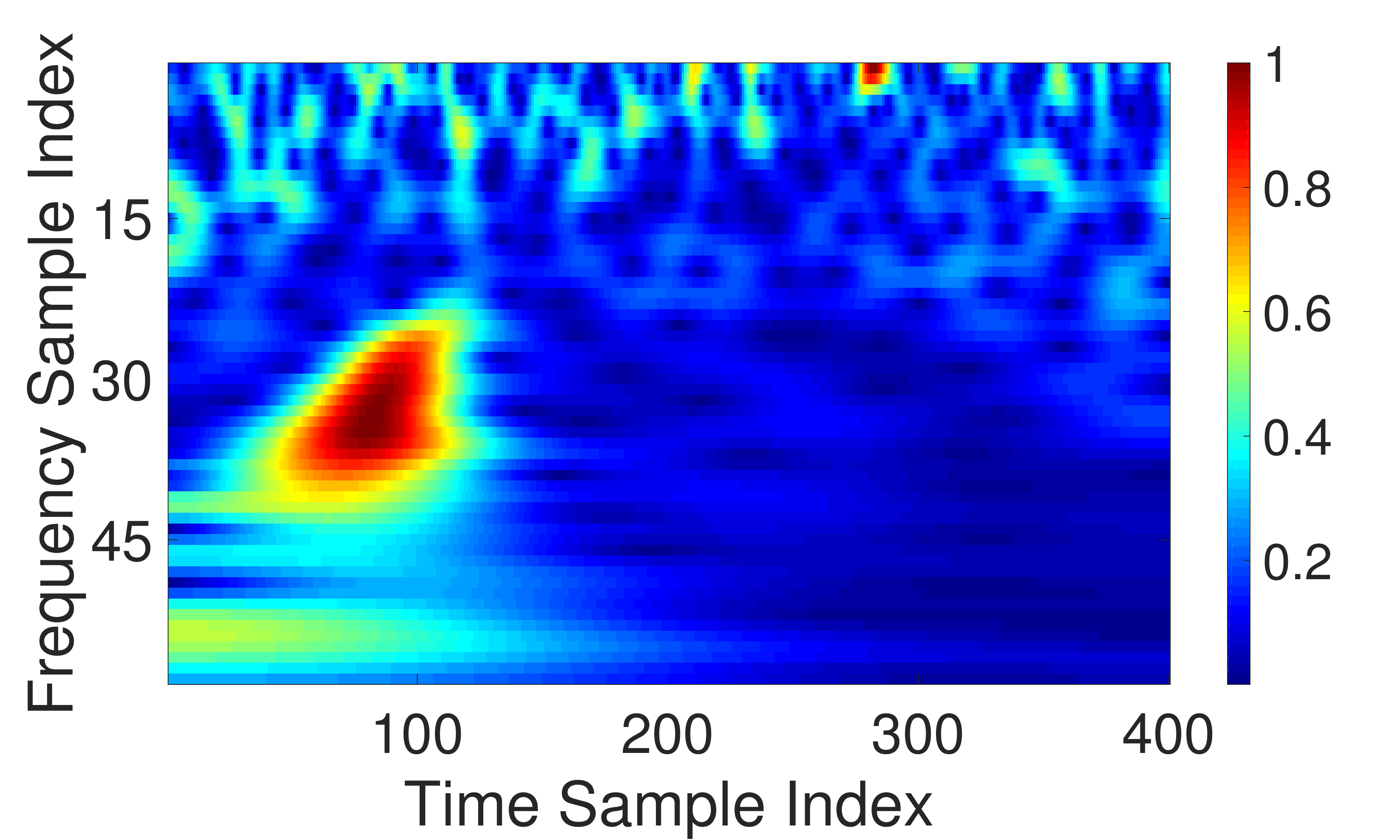}
	\label{fig:sit_uwb_cwt_2}	
}
\subfigure[Squat down: the 35-th fast-time index.]
{
	\centering
	\includegraphics[width=0.46\linewidth]{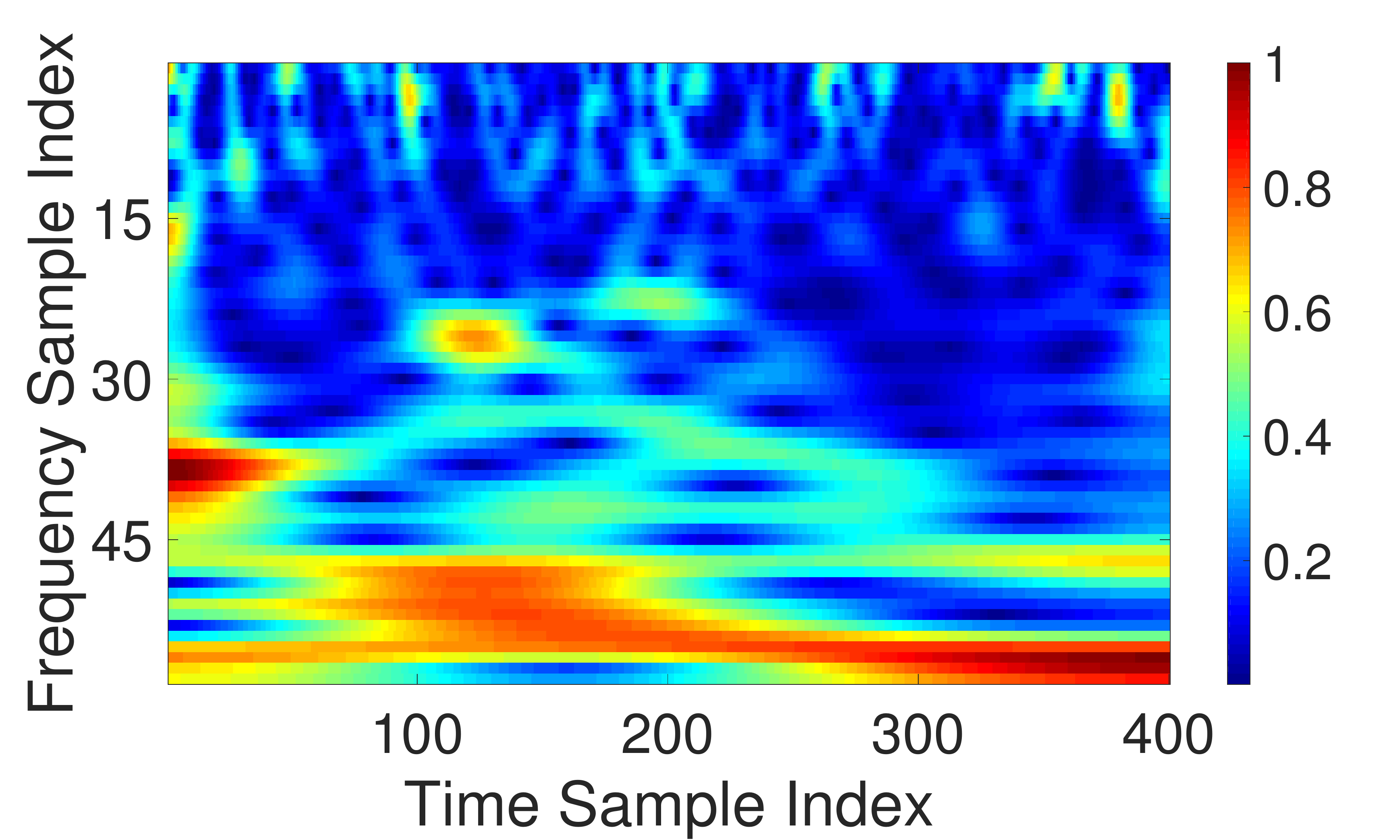}
	\label{fig:sqaut_uwb_cwt_2}	
}

	\caption{The time-frequency spectrograms of two similar activities obtained via WT based on our UWB radio. }
	\label{fig:uwb_act}
\end{figure}

To combat such limitations, we propose to employ UWB signals to obtain much richer features for more fine-grained sensing, and we demonstrate the power of UWB signals in Fig.~\ref{fig:uwb_act}. Owing to the large bandwidth that allows for sending very narrow pulses (which is impossible with the narrowband Wi-Fi), the motions of an activity can be ``sensed'' by multiple pulses.
We randomly select two fast-time indices of a pulse with each fast-time index containing 400 slow-time samples, and we employ the WT method on slow-time samples to obtain the time-frequency spectrogram. For more details of fast-time and slow-time, please refer to Sec.~\ref{sec:ssec:rf_chan}.
We can clearly see that, for the 22-nd fast-time index as shown in Fig.~\ref{fig:sit_uwb_cwt_1} and Fig.~\ref{fig:sqaut_uwb_cwt_1}, the shape of the hot zones are very different for sitting and squatting. This difference is further amplified for the 35-th fast-time index illustrated in Fig.~\ref{fig:sit_uwb_cwt_2} and Fig.~\ref{fig:sqaut_uwb_cwt_2}.
Essentially, the larger bandwidths we have, the richer and more distinctive features we can obtain to help classify activities more accurately.

\subsection{Crowded Channels}
Another practical issue with Wi-Fi-based sensing is that the Wi-Fi channels are usually very crowded~\cite{b51}. 
The accuracy of HAR is not only related to the proposed model, but also the quality of data.  If the recorded signal has lots of interference and noise, even though the model is very powerful, good performance can hardly be achieved. %

For existing Wi-Fi-based HAR systems, researchers usually control a dedicated Wi-Fi access point to send clean controlled Wi-Fi packets for HAR. This is not practical in real life because the controlled Wi-Fi packets occupy the precious  channel for data communication of the Wi-Fi AP. The uncontrolled Wi-Fi packets can hardly be used for HAR due to the random size, random time of arrival, and interference/noise from the surrounding Wi-Fi devices, bluetooth devices, and microwave appliances.  
Moreover, smart devices (e.g., smart speakers) also adopt Wi-Fi channels to transfer the contents for services. Although channel hopping can improve the signal quality, it may greatly affect the ongoing data communication~\cite{b40}. Therefore, it is safe to predict that Wi-Fi channels will become even more crowded in the future and they should not be competent candidates for HAR systems to achieve robust performance. %

\section{System Design} \label{sec:design}

\subsection{System Overview}
\begin{figure}[b]
	\centering
	\includegraphics[width=1\linewidth]{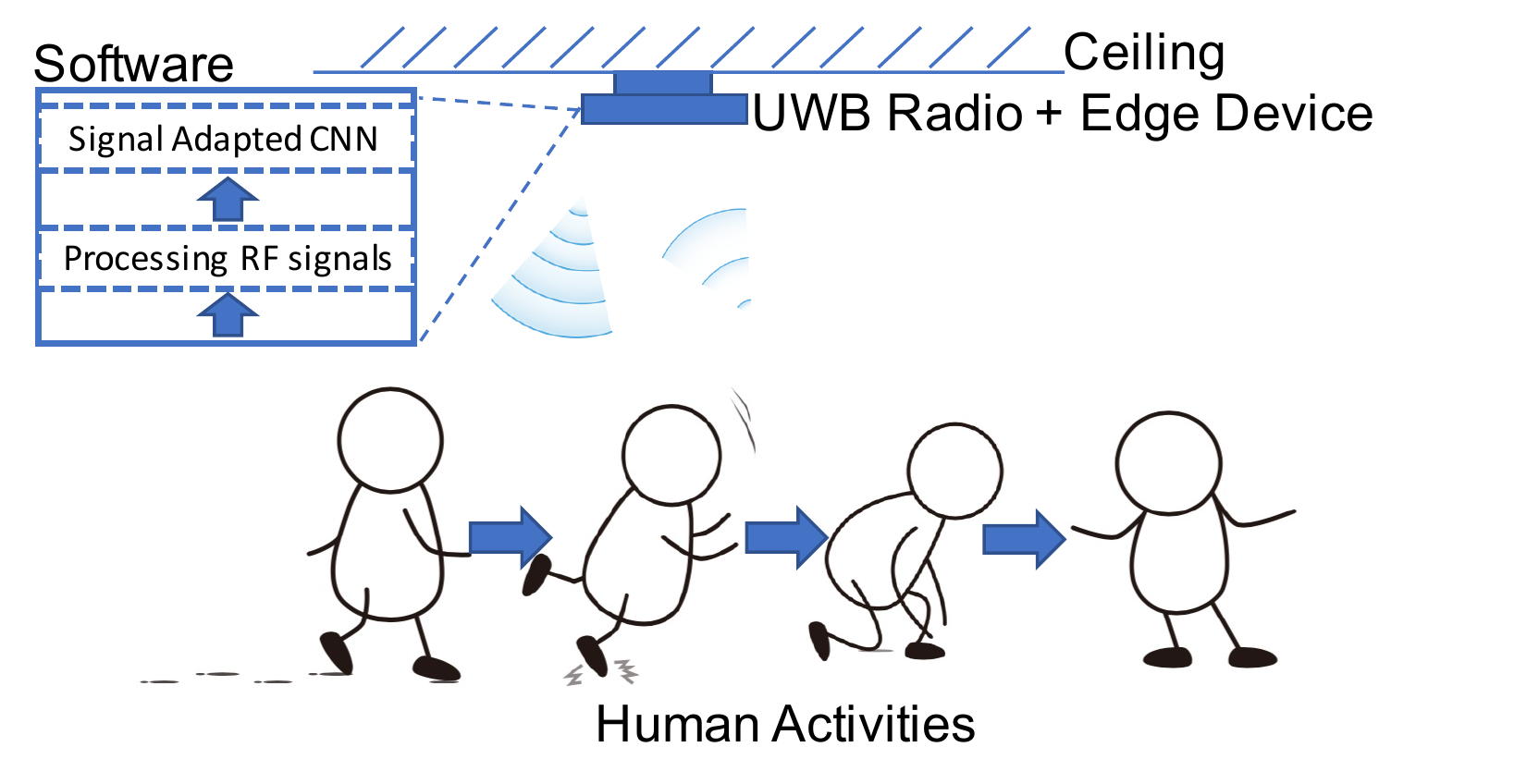}
	\caption{An overview of \systemname.}
	\label{fig:sys_overview}
\end{figure}

\systemname\ leverages  RF signals for passive HAR. It is built on a UWB radio and an edge device such as Raspberry Pi~\cite{b50} or ROCK Pi~\cite{rockpi}  as shown in Fig.~\ref{fig:sys_overview}.  Both the UWB  transmitter and receiver are collocated so it is convenient for them to be integrated into a single edge device. Note that, for Wi-Fi-based approaches, the transmitter and receiver are always two separated devices that are usually located at different locations. 
This integration also allows the edge device to directly control the UWB radio and to run the proposed algorithms for activity recognition. For software component, \systemname\ has two main algorithm modules.

\begin{itemize}
	\item \textbf{Signal Processing Module}: This module includes the denoising process and motion detection. After the reflections from the target are received by the UWB radio and delivered to the edge device, we employ a cascading filter to denoise the RF reflections.
	 Motion detection is designed to determine when the neural network model should be activated because non-activity samples may degrade the classifying performance of the model.
	\item \textbf{Signal Adapted CNN}: To the best of our knowledge, there is no CNN model design to accommodate both time and frequency domain information of RF signals for sensing. Therefore, we design a novel CNN structure to learn features from both time and frequency domains and to use these features for HAR. To realize a real-time activity recognition on edge device, a lightweight signal adapted CNN block is designed via employing efficient convolutions such as depth-wise dilated convolution, point-wise grouped convolution, etc. 
\end{itemize}
In the next few sections, we present the proposed RF channel model first, and then elaborate on each component of \systemname\ separately.

\subsection{Modeling RF Channel} \label{sec:ssec:rf_chan}
\begin{figure}[ht]
	\setlength\abovecaptionskip{-1pt}
	\setlength\belowcaptionskip{-1pt}
	\centering
	\includegraphics[width=1\linewidth]{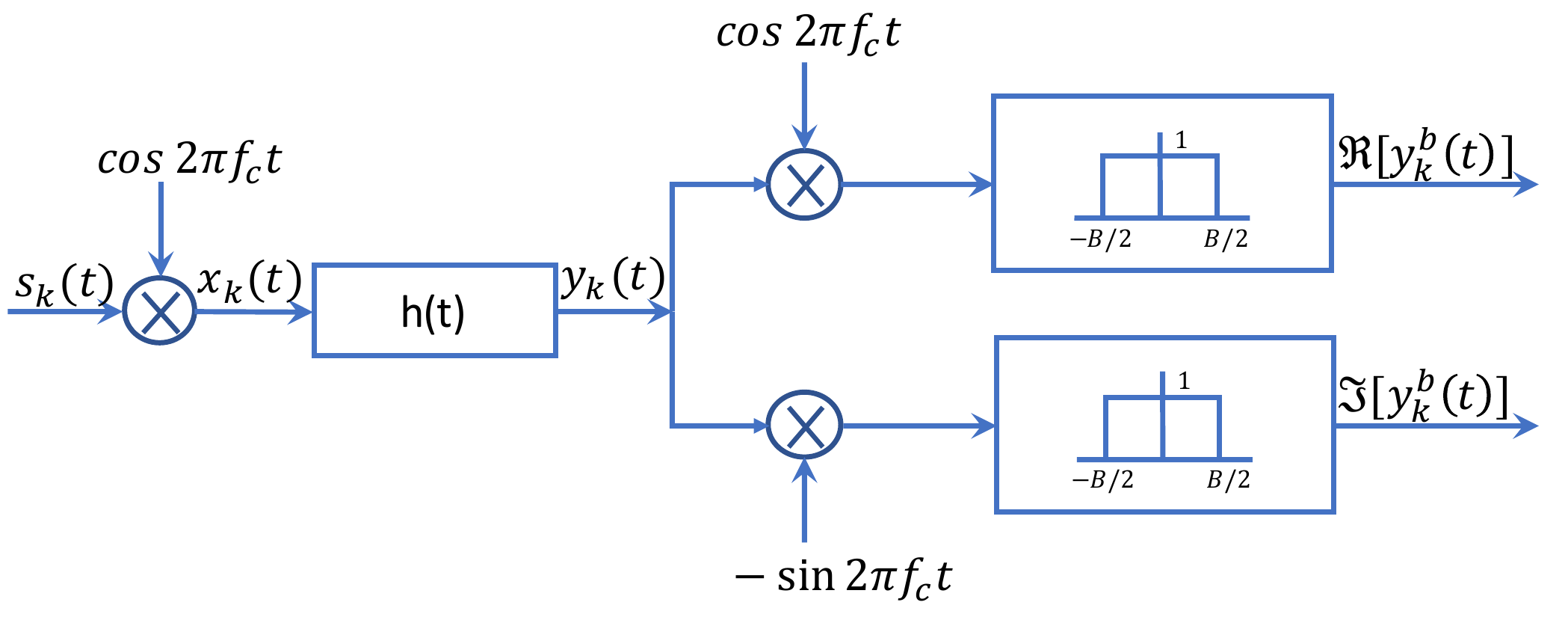}
	\caption{System diagram from the baseband transmitting 	signal $s_k(t)$ to the baseband
		received signal $y_k^b(t)$.}
	\label{fig:radarsysdiagram}
\end{figure}

In this section, we introduce the operations of UWB radio module. UWB impulse radio module works via transmitting pulse signal modulated by a carrier frequency. Note that ``pulse'' is loosely used, and the transmitted signal is not truly a pulse but has a very narrow width in the time domain. \systemname\ employs a commodity UWB radio module XETHRU \cite{b17} to capture RF signals reflected from targets. The system diagram of XETHRU from baseband transmitted signal $s_k(t)$ to received signal $y^b_k(t)$ is illustrated in Fig. \ref{fig:radarsysdiagram}. The radio architecture is different from typical  In-phase and Quadrature (IQ) sampling~\cite{b17}.  It only uses an in-phase single carrier frequency for upconversion, but IQ sampling at receiver for downconversion.  

\begin{figure}[h]
	\setlength\abovecaptionskip{-1pt}
	\setlength\belowcaptionskip{-1pt}
	\centering
	\subfigure[Time domain.]
	{
		\centering
\includegraphics[width=0.45\linewidth]{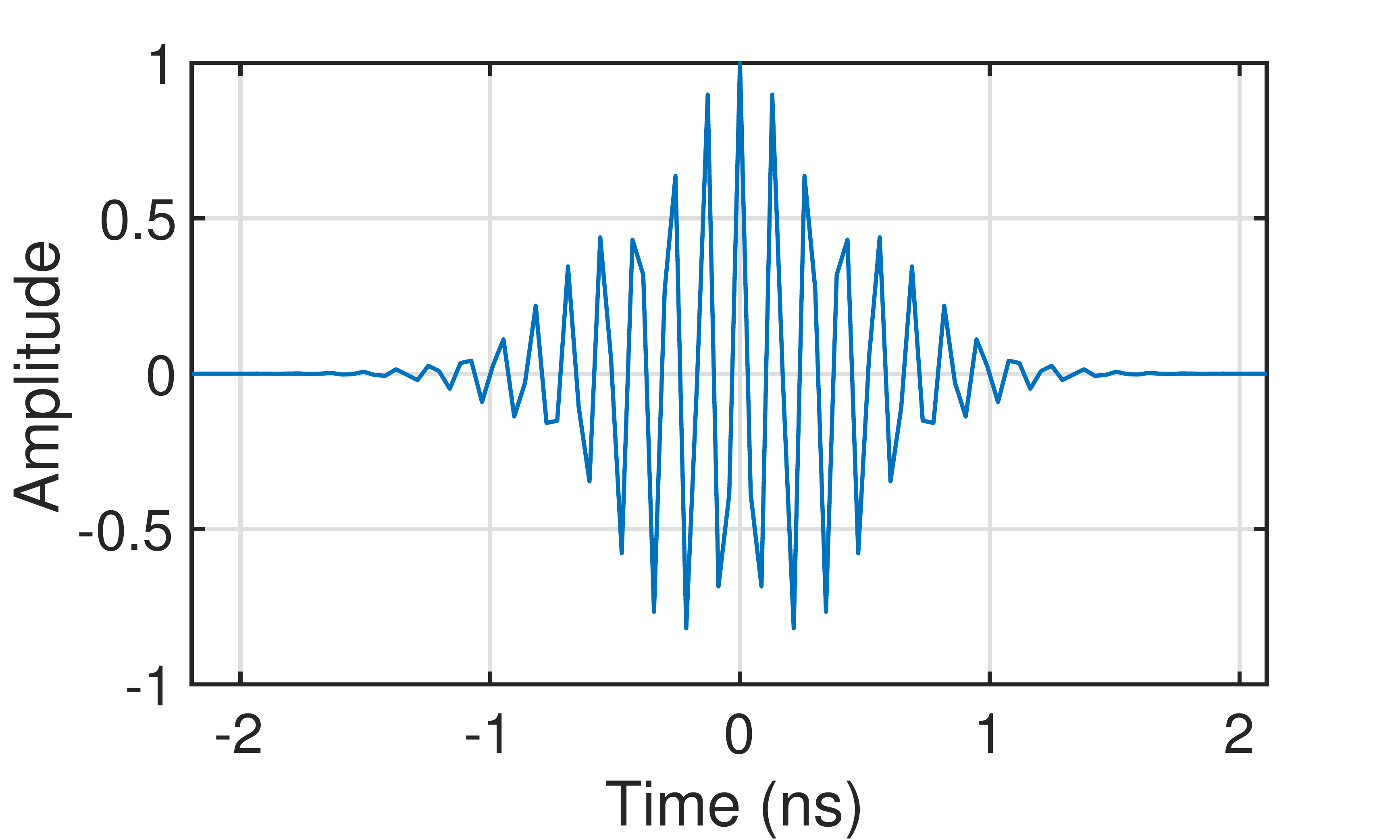}
\label{fig:uwbsig}
	}
	\subfigure[Frequency domain.]
	{
		\centering
\includegraphics[width=0.45\linewidth]{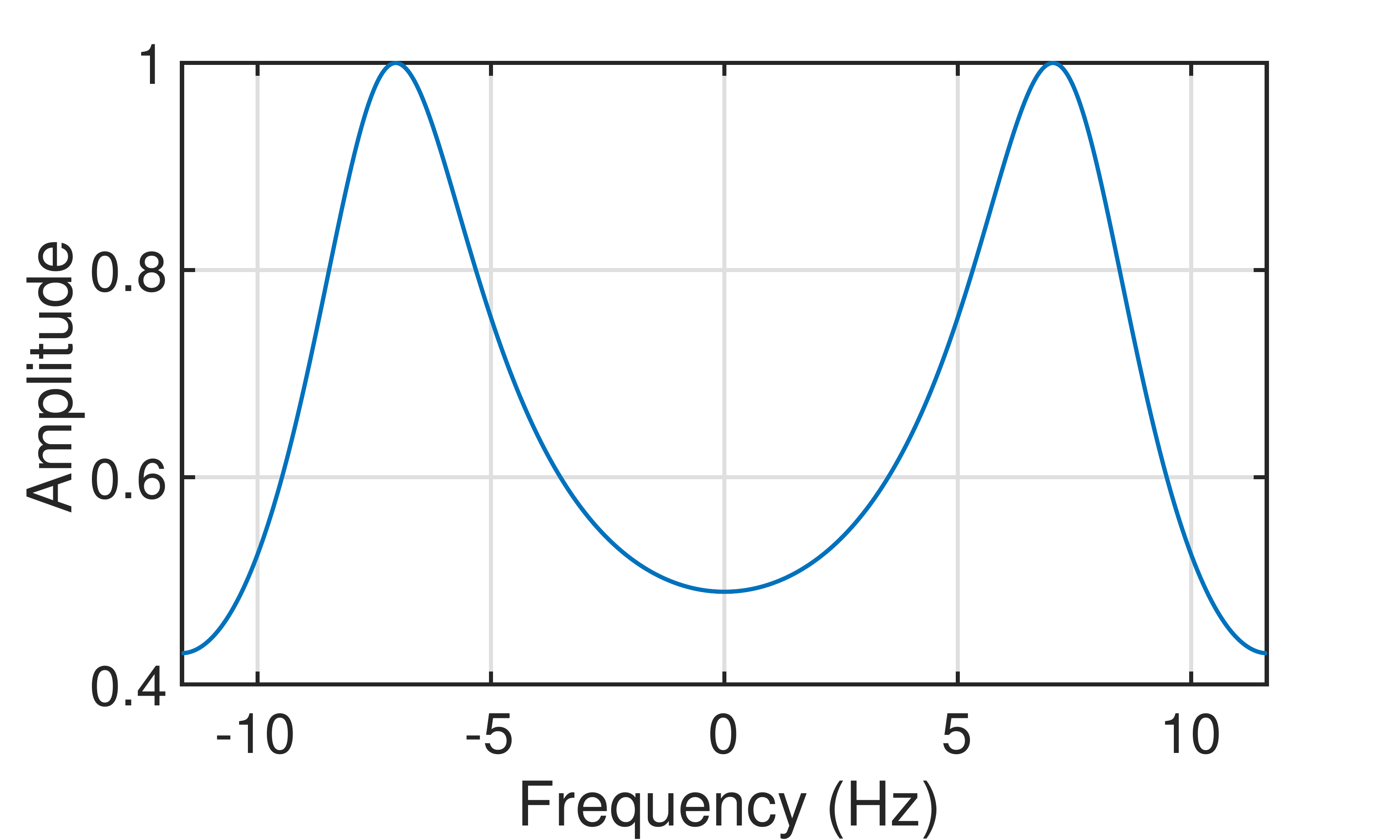}
\label{fig:uwbsigfre}
	}
	\caption{The transmitted signal $x_k(t)$.}
	\label{fig:uwb_tf}
\end{figure}

The transmitted signal, Gaussian pulse can be expressed as $s(t) = V_{tx}  \exp(- \frac{ (t - \frac{T_p}{2})^2 }{ 2 \sigma^2_p })$
where $V_{tx}$ is the pulse amplitude, $T_p$ is the signal duration, and $\sigma_p = \frac{ 1 }{ 2 \pi B_{-10dB} (\log_{10}(e))^{1/2} }$ is the \textcolor{black}{standard deviation that} determines the -10 dB bandwidth. After upconversion, transmitted signal in time domain at the $k$-th frame is given as
\begin{align} \label{eq:trans_sig}
x_k (t) = s(t - k T_s) \cdot \cos (2 \pi  f_c (t- kT_s ))
\end{align}
where $f_c$ is the carrier frequency, the operation $\cdot$ means a scalar product,  $T_s = \frac{1}{f_p}$ is the duration of the frame where $f_p$ is the pulse repetition frequency, and $s(t - k T_s) = s(t)$. For simplicity, we denote $t = t^{'} + k T_s$ with $t^{'} \in [0, T_s] $, and E.q \eqref{eq:trans_sig} can be written as $x_k (t) = s(t) \cdot \cos (2 \pi  f_c t).$
The transmitted signal $x_k(t)$ is illustrated in Fig. \ref{fig:uwbsig}, and its  frequency response is shown in Fig. \ref{fig:uwbsigfre}. The carrier frequency is 7.3 GHz, and bandwidth is 1.4 GHz. 

The Channel State Information (CSI) $h_k (t)$ with multipaths in a typical indoor environment is shown as following
\begin{align} \label{eq:multipath_csi}
h_k(t) = \sum_{p=1}^{P}  \alpha_p  \delta \left (t -  \tau_p    -  \tau_p^{D}(kT_s)   -  \tau_p^{mD}(kT_s)   \right )  
\end{align}
where $ \alpha_p $ is the propagation attenuation of the $p$-th reflection path signal,  $\tau_p$ is the time delay due to signal propagation, $\tau_p^{D}(k T_s)$ is the 
time delay caused by the large-scale body movement~(e.g., the human walking) and $\tau_p^{mD}(kT_s)$ represents the time delay caused by small-scale body movement~(e.g., the chest respiration movement). 
Moreover, for a transmitter-receiver collocated UWB radio,  $\tau_p = \frac{2 R_p}{c}  $, $\tau_p^{D}(kT_s) = \frac{ 2 v_p k T_s}{c}$, and $  \tau_p^{mD}(k T_s)   = \frac{ 2 \beta_p (1 - \cos (2 \pi \gamma_p kT_s))  }{c}  $ where $R_p$ is the distance between target and the UWB radio, $c$ is the signal propagation speed in the air, $v_p$ is the target movement speed, $\beta_p$ is the small-scale target displacement~(e.g., the chest displacement during respiration is around 0.5cm) which is usually smaller than one wavelength of radio wave and $\gamma_p$ is the movement frequency of target. Moreover, the range  resolution is inversely related to the channel bandwidth and is calculated with the following equation $\Delta r = \frac{c}{2B}$
where $B$ is the bandwidth of UWB radio. Hence, it is easy to calculate the  time delay resolution as $\Delta \tau = \frac{1}{2B}$. Thus, the received signals can be expressed as
\begin{align} \label{eq:radar_multipath_1}
y_k (t)  &= h_k(t) * x_k(t)  \cr
&= \sum_{p=1}^{P}  \alpha_p   \cos(2 \pi f_c (t - kT_s -  \tau_p    -  \tau_p^{D}(k T_s)   -  \tau_p^{mD}( k T_s)  )  \cr  
& \cdot s(t -  kT_s -  \tau_p    -  \tau_p^{D}(k T_s)   -  \tau_p^{mD}(k T_s)    )  + n(t)
\end{align}
where $n(t)$ is Gaussian noise with variance $\epsilon^2$ and the symbol $*$ is convolutional operation.  In practice, since $ kT_s \gg t $, the signal $y_k(t)$ is sampled in two dimensions: fast-time $t$ and slow-time $kT_s$.  The receiving baseband signals $y_k^b (t) $ are obtained after IQ downconversion.  We have 
\begin{align} \label{eq:radar_baseband}
y_k^b (t)  &=  \sum_{p=1}^{P}   \alpha_p  e^{ 2 \pi f_c (\tau_p  +  \tau_p^{D}(k T_s) + \tau_p^{mD}( k T_s) ) } \cr 
& \cdot s(t -  kT_s -  \tau_p    -  \tau_p^{D}(k T_s)   -  \tau_p^{mD}(k T_s)    )  + n(t).
\end{align}
Different human activities exhibit different $\tau_p$, $ \tau_p^{D}(k T_s)$, and $ \tau_p^{mD}(k T_s)   $ in  $y_k^b(t)$. Therefore, the received UWB signal contains richer 
features for HAR. 

\begin{figure}[ht]
		\centering
		\includegraphics[width=0.5\linewidth]{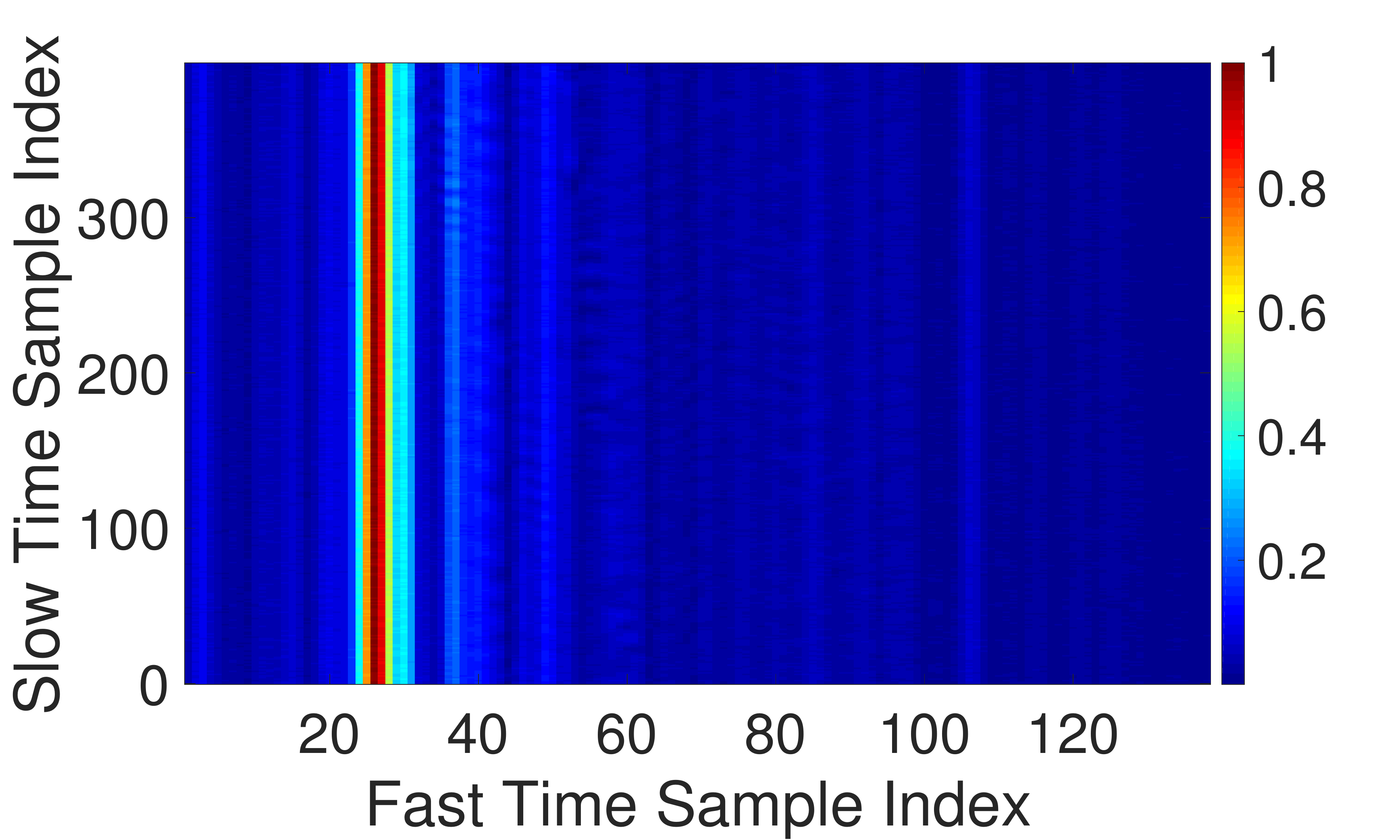}
		\caption{The matrix of receiving baseband signals.}
		\label{fig:fast_slow_time}
\end{figure}

Let $t = lT_{n}$ represent the $l$-th discrete  sample via Analog to Digital Conversion~(ADC) where $T_n$ is the sampling interval. Thus, the discrete baseband signals are $y_k^b (lT_n) $.  The received signals can be formed as a matrix along with fast-time and slow-time shown in Fig.~\ref{fig:fast_slow_time}. 
In general, fast-time axis indicates time delays caused by range distance, and the slow-time axis is used to estimate Doppler information via a long time window observation.

\subsection{Processing RF Signals}
Before feeding data into the neural network,  the noise caused by hardware and environment needs to be removed to \textcolor{black}{enhance signal quality}. 
In addition,  since our CNN model is proposed for HAR, signal samples in the non-activity scenario are removed from the training and inference stages; otherwise such samples may introduce errors in classification. Consequently, the RF signal processing has three main steps: i) phase noise reducing, ii) signal SNR enhancement, and iii) motion detection. 

\subsubsection{Phase Noise Reducing}
\begin{figure}[h]
	\centering
	\subfigure[The raw signals.]
	{
		\centering
		\includegraphics[width=0.45\linewidth]{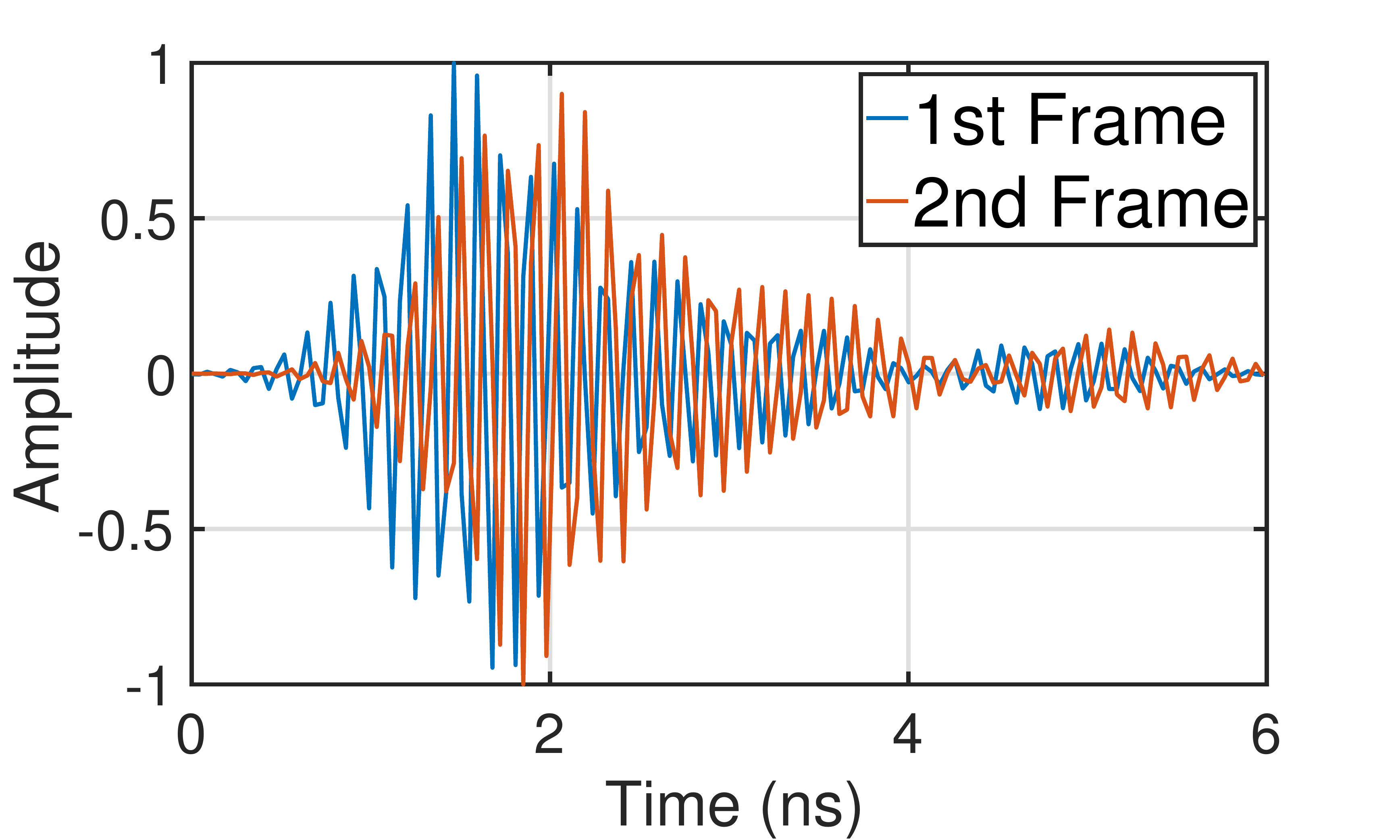}
		\label{fig:pn_example}
	}
	\subfigure[The baseband signals]
	{
		\centering
		\includegraphics[width=0.45\linewidth]{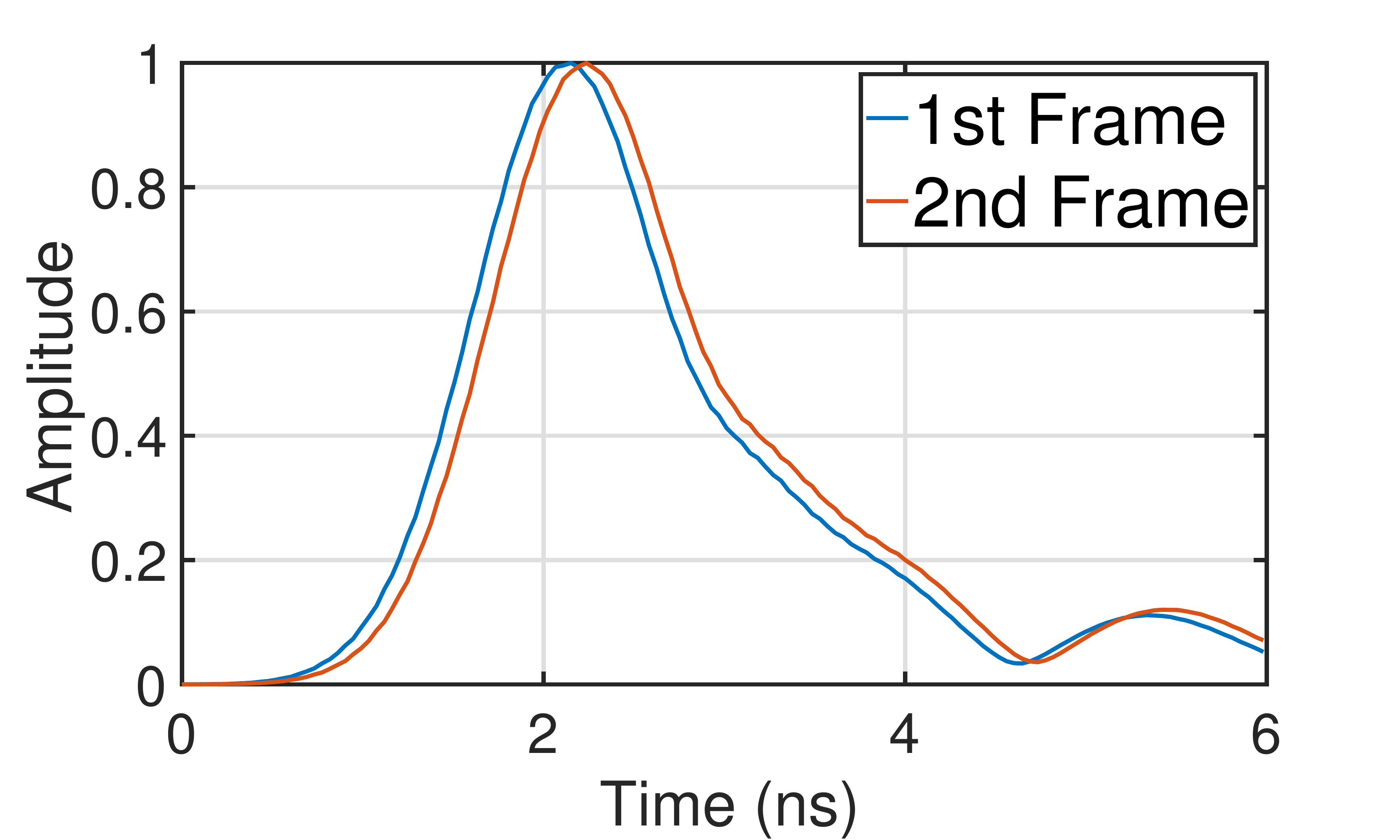}
		\label{fig:np_example_bb}
	}
	\caption{Signals with phase noise.}
	\label{fig:np}
\end{figure}

The ADC of UWB signals introduces Sampling Timing Offset~(STO) caused by imperfect sampling clock. The signal phase perturbed by such STO will affect the Doppler and Micro Doppler information. 
Doppler and Micro Doppler are observed via slow-time $kT_s$. If the reflection is from a static object, the phase introduced by both Doppler and Micro Doppler is zero.
In Fig.~\ref{fig:pn_example},  two subsequent raw RF signals frames (slow-time) $y_k(t)$ with phase noise caused by STO are reflected from a same static object. The baseband signal after IQ downconversion $y_k^b(t)$ with two subsequent frames are shown in Fig. \ref{fig:np_example_bb}.  We can see that the amplitude of the second frame with phase noise has jitter, but ideally, the amplitudes of two frames should be the same.  For HAR,  the static objects may be considered as moving because of this phase noise. 

\begin{figure}[b]
	\centering
	\includegraphics[width=0.7\linewidth]{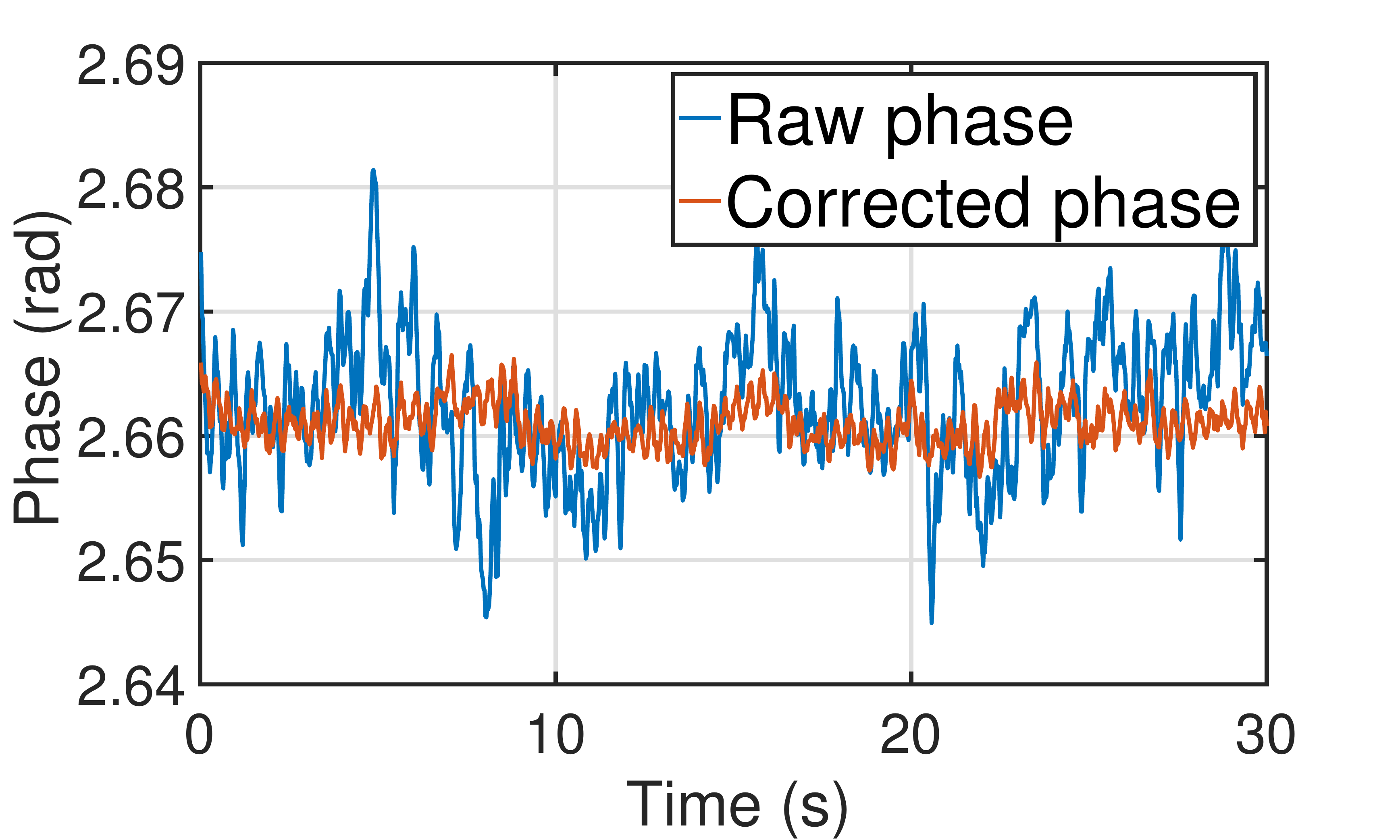}
	\caption{Phase with noise correction.}
	\label{fig:phasenoise}
\end{figure}

{\color{black} The phase of an object with jitter is $\Omega_p +  \Delta \omega(t - kT_s) $.  The phase jitter caused by STO at the radio receiver is $\Delta \omega(t - kT_s)$, and the phase of signal reflected from a object is  $  \Omega_p $. Our objective is to reduce the phase jitter $\Delta \omega(t - kT_s)$. }
To achieve this objective, firstly, we need to find a reflector signal from a static object as the reference. For instance, we can choose the pulse with maximum amplitude.  Secondly, for the $K$ frames, we can calculate the mean phase  $ \hat{\omega} $ of that pulse. Then, we calculate the difference between $\hat{\omega}$ and phase of the reference signal at the $k$-th frame. Finally, we adjust the phases of all samples in fast-time with the above difference. %
We can see that the phase with time domain noise correction is much more stable as shown in Fig. \ref{fig:phasenoise}. 
In practice, since our system is mounted on the ceiling, the max peak of reflection is always the floor with the largest Radar Cross Section~(RCS). Therefore, one may readily identify such reflections out of those from other static objects to correct the phase. As the variation in phase is very small, it barely affects the inference results of our neural network, due to the large motions of human activities. However, this variation could make the training phase unstable, resulting in a longer convergence time.

\subsubsection{ SNR Enhancement }
\begin{figure}[h]
	\centering
	\subfigure[The received signal without SNR enhancement.]
	{
		\centering
		\includegraphics[width=0.45\linewidth]{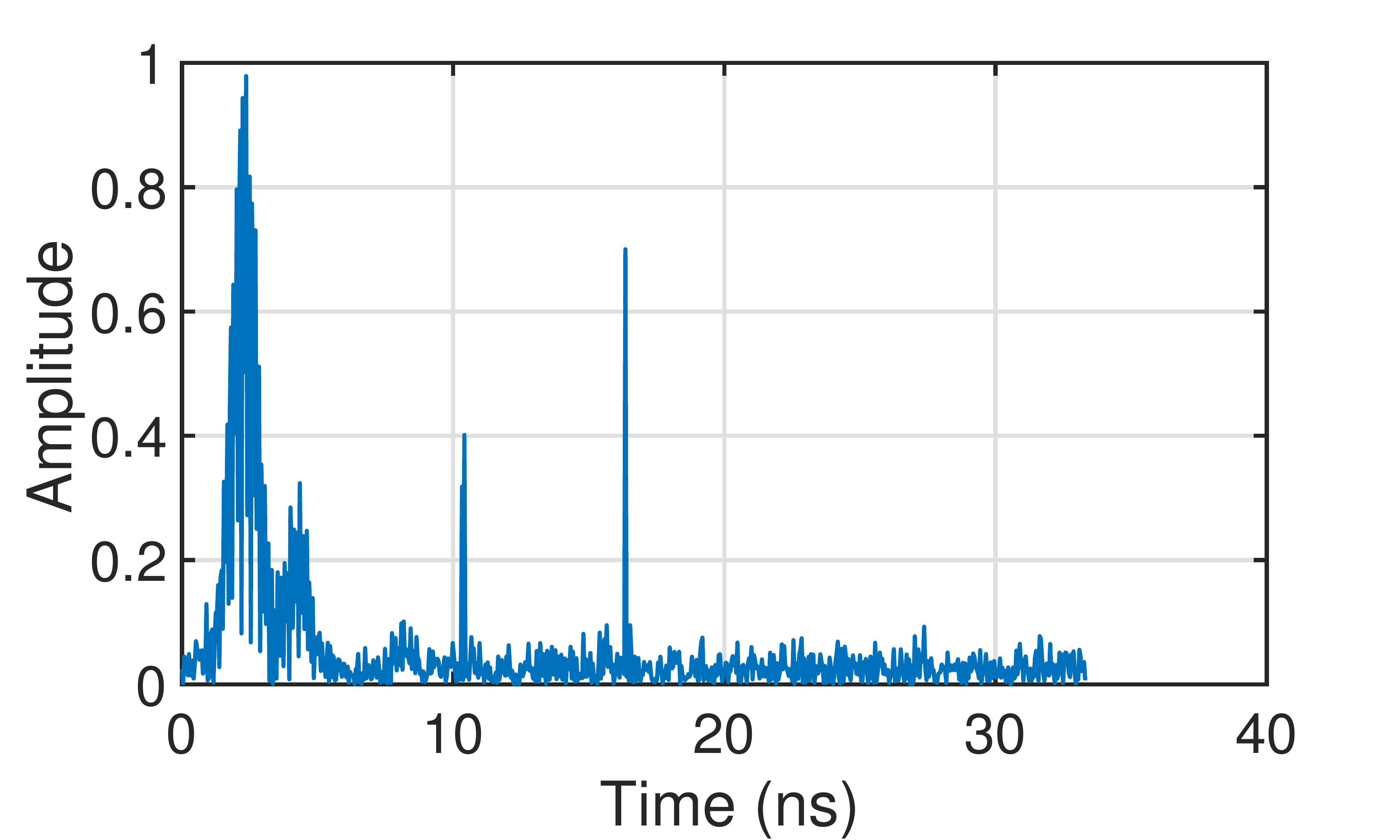}
		\label{fig:nonfilter}
	}
	\subfigure[The received signal with SNR enhancement.]
	{
		\centering
		\includegraphics[width=0.45\linewidth]{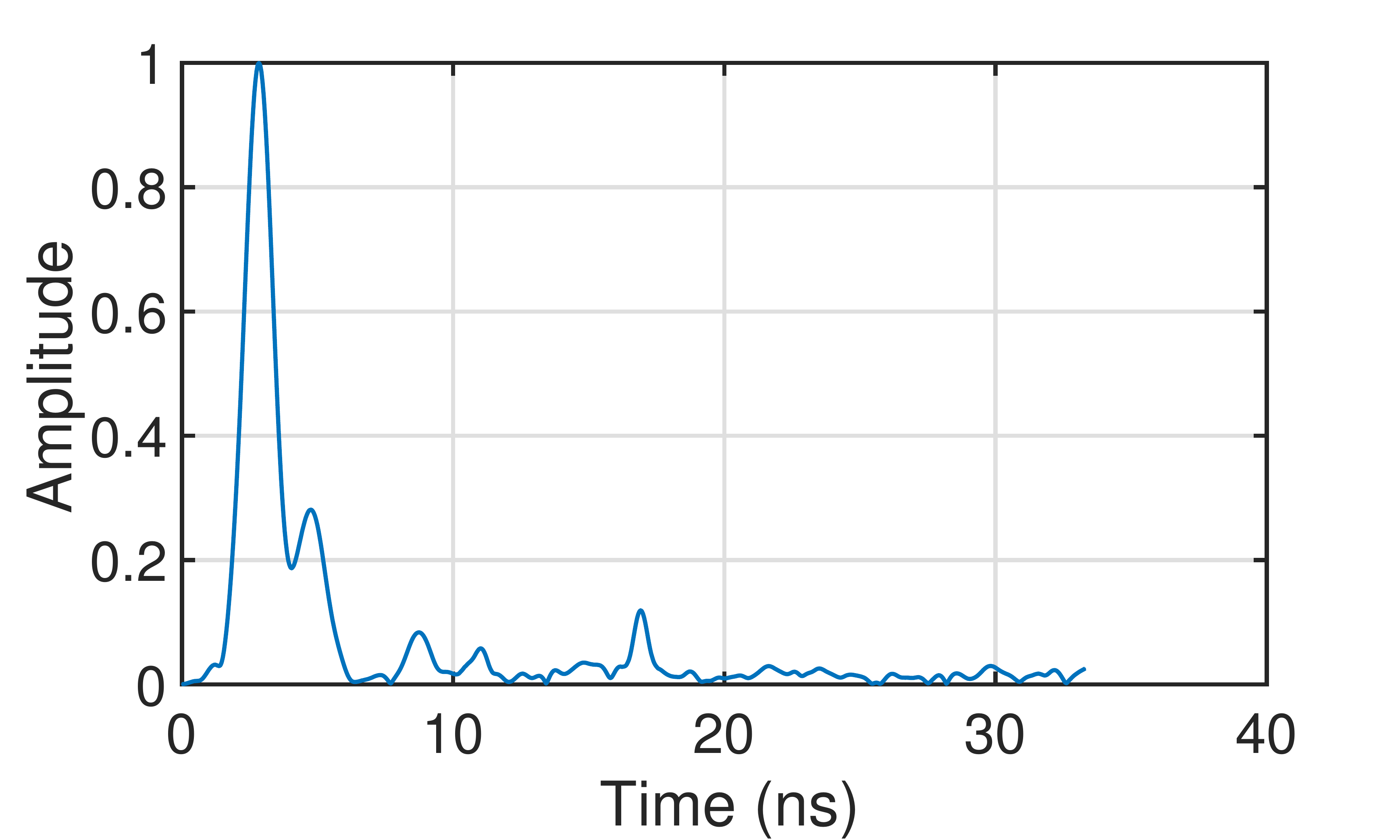}
		\label{fig:filter}
	}
	\caption{SNR enhancement can improve the quality of signals.}
	\label{fig:wofilter}
\end{figure}

The raw receiving baseband signals are corrupted by noise as shown in Fig. \ref{fig:nonfilter}. The noise  brings in errors in the neural network model.  Specifically, if not properly addressed, the random noise will be learned by the neural network model that tends to overfitting. Consequently, we leverage a cascading filter to remove noise and enhance the SNR of the received baseband signal. The cascading filter includes a low-pass filter and a smoothing filter. We first adopt a Finite Impulse Response (FIR) low pass filter with 26 taps and a hamming window. %
Then a smoothing filter with 5-point window is applied to smooth the output from the FIR low pass filter. Fig.~\ref{fig:filter} illustrates the signal output after cascading filter, showing noise being greatly suppressed. 

\subsubsection{ Motion Detection}

\textcolor{black}{Before feeding data to the classifier, we face two practical
issues: i) detect the human motion within a certain range and ii) identify the starting point} of a human activity. Fortunately, we observe that, due to the high temporal resolution, a human activity naturally spans several fast-time samples \textcolor{black}{and the peak power indicates a motion after removing the static background reflections. As a result, the peak power enables us to detect the human motion on one hand, while its fast-time index
also signals the start 
of a human activity on the other hand.} Next, we explain the design principle for motion detection module.

\begin{figure}[b]
	\centering
	\begin{minipage}[t]{0.48\linewidth}
		\centering
		\includegraphics[width=1\textwidth]{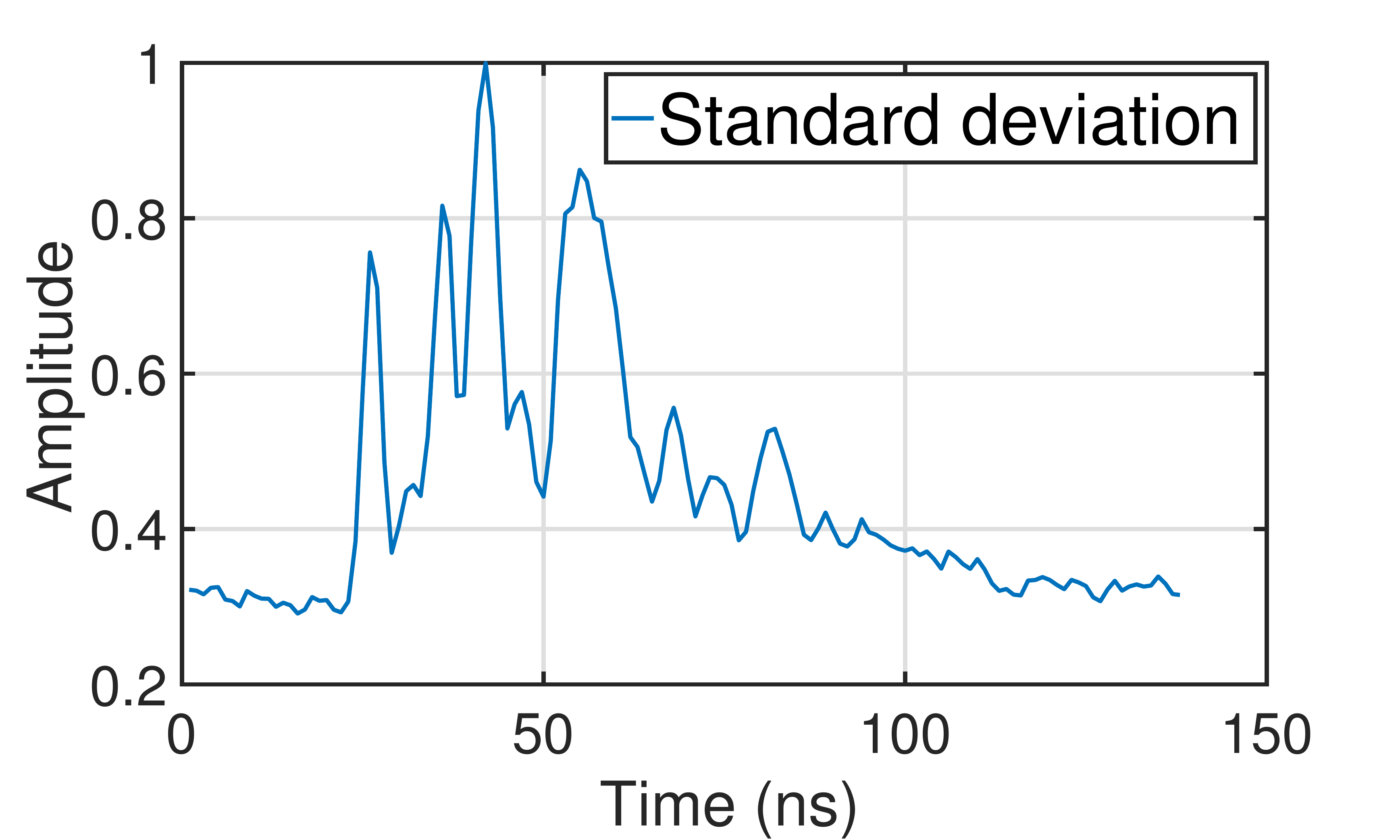}
		\caption{The standard deviation of UWB radio signals.}
		\label{fig:cfar1}
		\hspace{-0.05cm}
	\end{minipage}
	\begin{minipage}[t]{0.48\linewidth}
		\centering
		\includegraphics[width=1\textwidth]{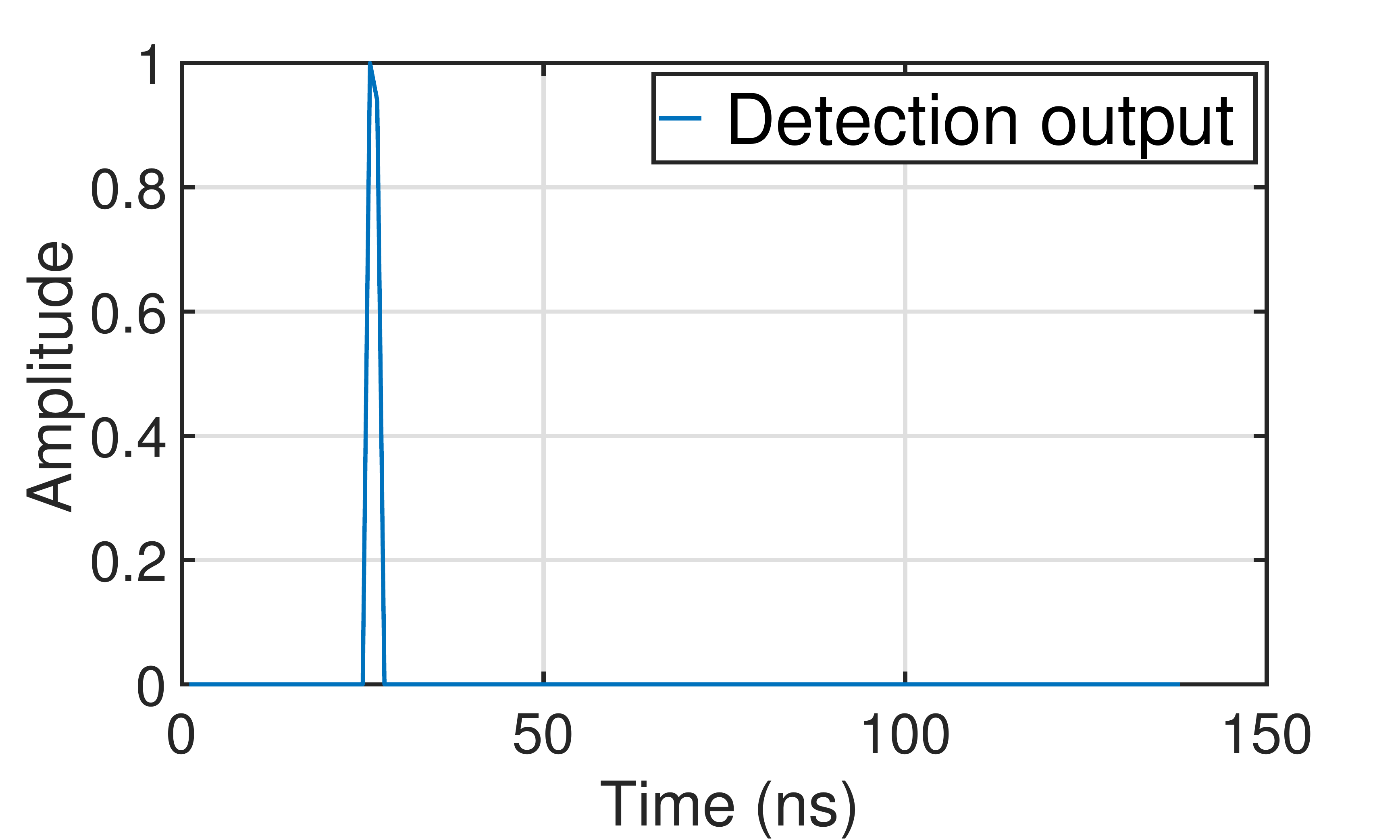}
		\caption{Motion detection output.}
		\label{fig:cfar2}
	\end{minipage}
\end{figure}

We remove the static environment via background subtraction \cite{b52}. 
The standard deviation and peak-average detection algorithm are employed to detect human motions. The standard deviation $ SD$ is calculated as $ \sqrt{ \sum_{n = 1}^{N}( v_i - \bar{v} )^2 /(N - 1 ) }$ where $\{ v_1, v_2, \cdots, v_N \}$ are the observed values. %
For $l$-th pulse in fast-time, we calculate the standard deviation.  
The standard deviation vectors $\{SD_1, SD_2, \cdots, SD_L \}$ are obtained as shown in Fig.~\ref{fig:cfar1}.  There are multiple peaks in Fig.~\ref{fig:cfar1}, and each peak indicates one movement in the environment. The key insight is that the wireless signals are impacted by objects in motion, and the standard deviation of each sample in fast-time is larger when there are motions. 
\begin{figure}[t]
	\centering
	\includegraphics[width=0.7\linewidth]{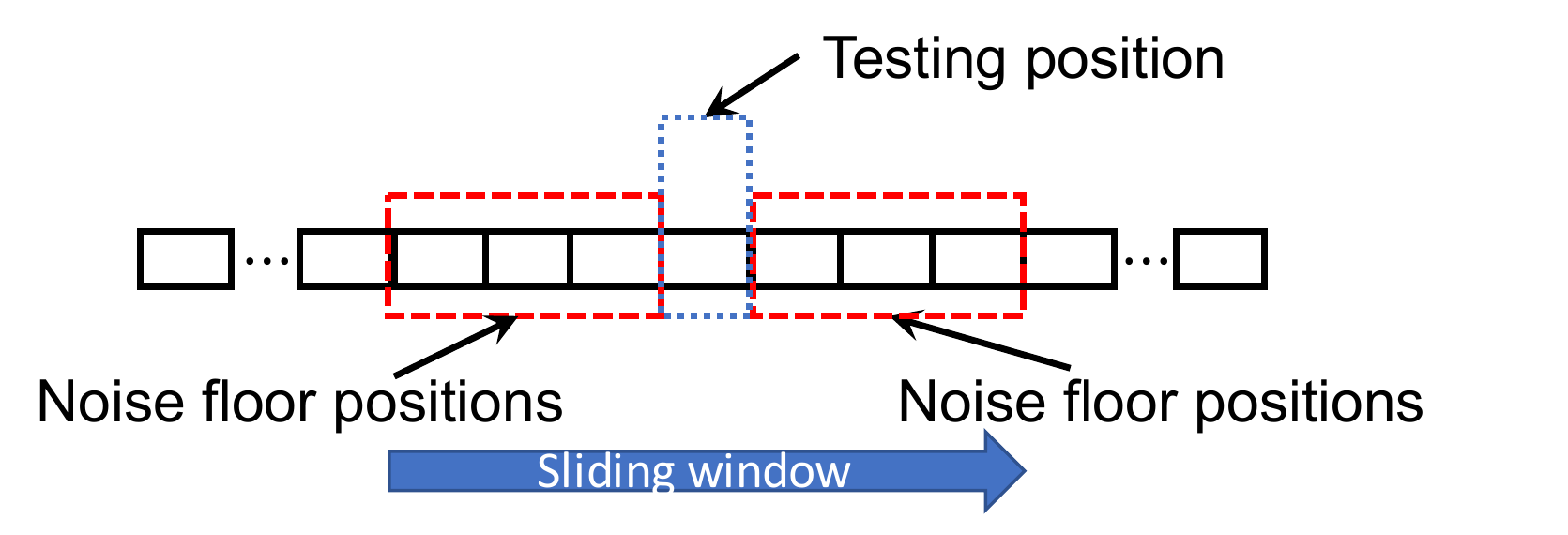}
	\caption{The procedure of peak-average detection algorithm.}
	\label{fig:cfar3}
\end{figure}
However, not only human create motions, but also electronic fans, air conditioners, pets, etc. Consequently, we design the peak-average detection algorithm to avoid false alarms. %
Since the noise level changes both spatially and temporally, we cannot use a fixed threshold to detect human motion. Luckily, there is an observation that human targets always perform larger activities than the interferers.  Consequently,  the larger standard deviation in the position is human motion.
The process of the algorithm is illustrated in Fig.~\ref{fig:cfar3}. It is clear to see that the noise floor threshold  $th_\mathrm{motion}$ can be estimated by averaging the values at all noise floor positions, and the value $val$ at a testing position is compared with $ coef \cdot th_\mathrm{motion}$where $coef$ is a constant to adjust the threshold. 
Empirically, we choose $coef = 1.5$ in our design. The detection output of data used in Fig.~\ref{fig:cfar1} is shown in Fig. \ref{fig:cfar2} which indicates a human motion is detected. %

\subsection{ Signal Adapted  Convolutional Neural Network }
The conventional CNN is designed for computer vision \cite{b16}. Although many researchers directly apply  such neural network to extract features for wireless sensing applications \cite{b6,b9,b10,b12,b20},  they only utilize features in time or frequency domain~\cite{b19}. Most Wi-Fi based approaches  employ frequency domain information because Wi-Fi signals contain less time domain information due to low time resolution. 
Recently, a few works such as~\cite{b53} consider both time and frequency domain information to design HAR systems, but their system are not an end-to-end learning system. They only use neural network to deal with frequency domain, and extract features from time domain in a handcarft method. 
More importantly, the previous works usually need a powerful computer to run the CNN model and infer the activities. Such computation-heavy CNNs have difficulties to be run on resource-constrained edge devices. We thus propose a signal adapted CNN to address the above two challenges in this section. 

We take the unique property of RF signals into consideration to design a lightweight CNN. The dominant reflections come from different body parts as the person moves over time. For instance, different from camera, at each time slot, the received signals only reflect back from a subset of body parts. To deal with these issues, we make our CNN model aggregate information from both time and frequency domains to extract the features for different activities. 
In our design, one slow-time frame contains a total of 60 fast-time samples, which are the delay profile of a single pulse. The slow-time frames are sampled at 400\!~Hz; they are respective signal delay profiles of consecutive pulses.
Consequently, a  $400 \times 60 $ spectrogram in time domain is taken as input.  To obtain frequency domain information of signals, we perform Fast Fourier Transform~(FFT) instead of WT.  The reason is that if we perform WT on each fast-time index, the data size is $400 \times 60 \times W $, where $W$ is the number of wavelet series, usually in the scale of 50-60. Such a large data size will incur a huge computation overhead.

\begin{figure}[b]
	\centering
	\includegraphics[width=0.6\linewidth]{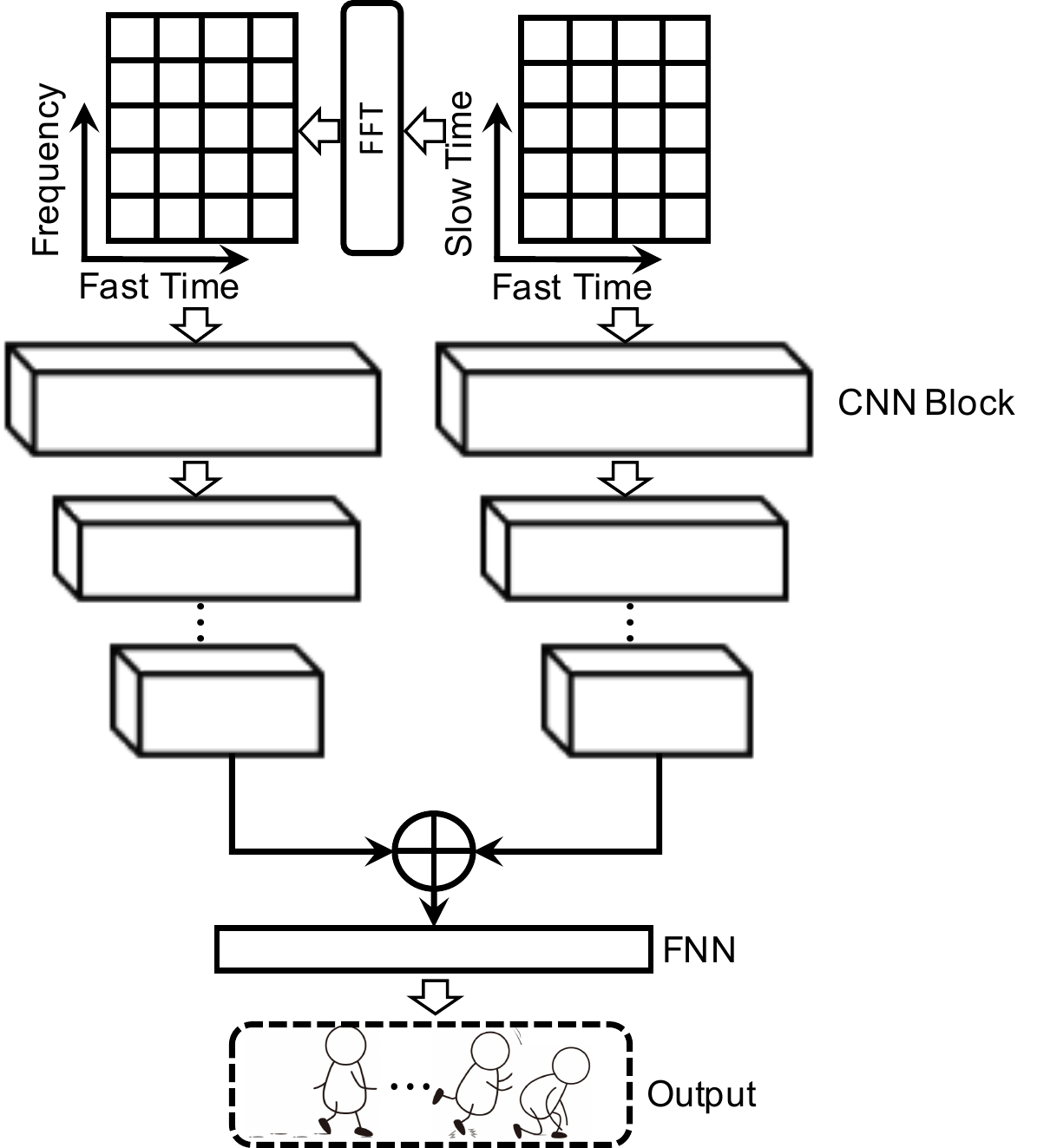}
	\caption{Our convolutional neural network design.}
	\label{fig:cnndesign}
\end{figure}

Our signal adapted CNN is a two-stream (time and frequency) CNN architecture composed of two parts. 
The simplest way to fuse two spectrograms of time and frequency is to put those spectrograms into different two channels of an image, and then we can feed them to the CNN. %
However, in this way, the time and frequency spectrograms will correspond to different pixels in the image. %
Therefore, in our design, we use separated branches to extract features from spectrograms of time and frequency, respectively as shown in Fig.~\ref{fig:cnndesign}. In this architecture, each branch does not share the CNN layer weights with the other.
Each branch has multiple efficient CNN blocks to abstract high level features, \textcolor{black}{and three blocks are adopted in \systemname}.  Each CNN block output is followed by an activation function ReLU that is computed via  the function $f(x)=\max(0,x)$.  
Then, two branches are aggregated via a concatenation operation $ \oplus $ and put into fully connected layers $ f_{\text{FNN}}(T \oplus F) $ where the symbol $T$ represents the features of time, and the symbol $F$ illustrates  that of frequency. Finally, a softmax function shown as following is employed to achieve probability prediction for each class
$
f_i^{\text{sm}}(\bm{X}) = \frac{e^{x_i}}{\sum_{j=1}^{K} e^{x_j} }   
$, where $\bm{X} = \{ x_1, \cdots, x_K \} $ is the input vector. In the end, the input vector is normalized by the sum of all exponential functions. Finally, we use the cross entropy $L = - \sum_{c=1}^{C} y_c \log(p_c)$ as the loss function in our system where $C$ is the total number of classes and $p_c$ is the probability of the $c$-th class.

For resource-constrained edge device, we need to implement multiple efficient CNN blocks to build the above signal adapted model. To build an efficient model,  we resort to the power of convolution factorization. The key idea is to employ a factorized version such as depth-wise separable convolution which consists of depth-wise convolution and point-wise convolution~\cite{b41}  or group convolution~\cite{b42} to replace the traditional full convolutional operation.  We assume a standard convolution operation with an input $\bm{ X } \in \mathcal{R}^{W \times H \times c_{in}}$, a convolutional kernel $\bm{K} \in \mathcal{ R }^{k \times k \times c_{in} \times c_{out} }$, and an output $ \bm{Y} \in \mathcal{ R }^{ W \times H \times c_{out} } $. 
For each output of a filter $\bm{W}$, the mathematical formulation of traditional CNN, point-wise convolution, depth-wise convolution and group convolution, respectively are  
\begin{align} \label{eq:cnn_tradition}
\text{Conv}( \bm{W}  , \bm{ X })_{(i,j)} &= \sum_{l = 1}^{ k}  \sum_{m = 1}^{ k}   \sum_{c=1}^{c_{in}} \bm{W}_{(i, j ,c )} \cdot \bm{X}_{(i+l,j+m,c)} 
\end{align}
\begin{small}
\begin{align} \label{eq:cnn_pointwise}
\text{PConv}( \bm{W}  , \bm{ X })_{(i,j)} &=  \sum_{c=1}^{c_{in}} \bm{W}_{c} \cdot \bm{X}_{(i,j,c)} 
\end{align}
\end{small}
\begin{small}
\begin{align} \label{eq:cnn_depthwise}
\text{DConv}( \bm{W}  , \bm{ X })_{(i,j)} &= \sum_{l = 1}^{ k}  \sum_{m = 1}^{ k}    \bm{W}_{(i, j ,c )} \cdot \bm{X}_{(i+l,j+m,c)} 
\end{align}
\end{small}
\begin{small}
\begin{align} \label{eq:cnn_group}
\text{GConv}( \bm{W}  , \bm{ X })_{(i,j)} &= \sum_{l = 1}^{ k}  \sum_{m = 1}^{ k}  \sum_{c=1}^{c_{in}/G}    \bm{W}_{(i, j ,c )} \cdot \bm{X}_{(i+l,j+m,c)}. 
\end{align}
\end{small}
Moreover, the depth-wise separable convolution is
\begin{small}
\begin{align} \label{eq:cnn_sep}
\text{SConv}(\bm{W}_p,\bm{W}_d ,  \bm{X})_{i,j} &= \text{PConv}_{i,j}(\bm{W}_p, \text{DConv}_{(i,j)  } ( \bm{W}_d, \bm{X} ) ). 
\end{align}
\end{small}
According to Eq.~\eqref{eq:cnn_tradition}, we realize that for each filter, the size of effective receptive field is $k \times k $,
and the number of learning parameters are $ k^2 c_{in} $.  For a number of $c_{out}$ filters, we have a total of $k^2 c_{in} c_{out}$ parameters for the convolutional kernel. Also, point-wise convolution in Eq.~\eqref{eq:cnn_pointwise} and depth-wise convolution in Eq.~\eqref{eq:cnn_depthwise} show that the total number of parameters are $c_{in} c_{out}$ and  $k^2c_{in} $, respectively. Therefore, when we use depth-wise separable convolutional operation,  according to Eq.~\eqref{eq:cnn_sep}, the number of parameters is significantly decreased to $ k^2c_{in} + c_{in} c_{out}$~\cite{b41}, by smartly combining depth-wise and point-wise convolutions.

\begin{figure}[t]
	\centering
	\includegraphics[width=0.3\linewidth]{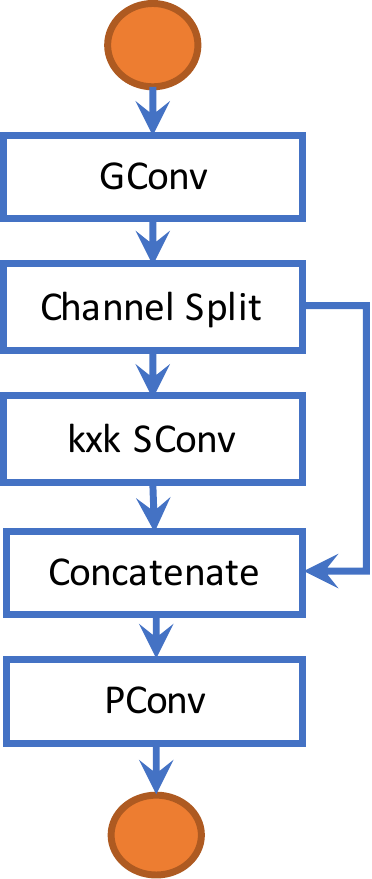}
	\caption{The CNN block in \systemname. GConv divides different sets of channels to perform convolution operation independently; these independent group computing can be paralleled. Channel split separates input channels equally to each SConv.}
	\label{fig:cnnblock1}
\end{figure}
Our efficient block architecture is illustrated in Fig.~\ref{fig:cnnblock1} based on a reduce-split-transform-merge rule.  We use a $1\times 1$ group convolution to reduce the number of parameters of channels from $c_{in} c_{out} $ to $\frac{c_{in}  c_{out} }{G}$. \textcolor{black}{There are 3 layers in each CNN block.} Note that different groups can be computed in parallel. To reduce the amount of computations, the channel split module divides the input features into two branches. One branch is applied with the $k \times k$ depth-wise separable convolution, and the other is concatenated with the output of the first branch.
Finally, we use a point-wise convolution to enable the information communication among different channels. The features of wireless signals are not like the camera images, and they are sparse in the spectrograms. Hence, the dilated convolution~\cite{b43} is applied in our network block to enable large effective receptive fields.

\section{Implementation and Evaluation} \label{sec:impl_eval}

\subsection{Implementation}\label{sec:impl}
\begin{figure*} [ht]
	\centering
	\begin{minipage}[t]{0.33\linewidth}
		\centering
		\includegraphics[width=.68\linewidth]{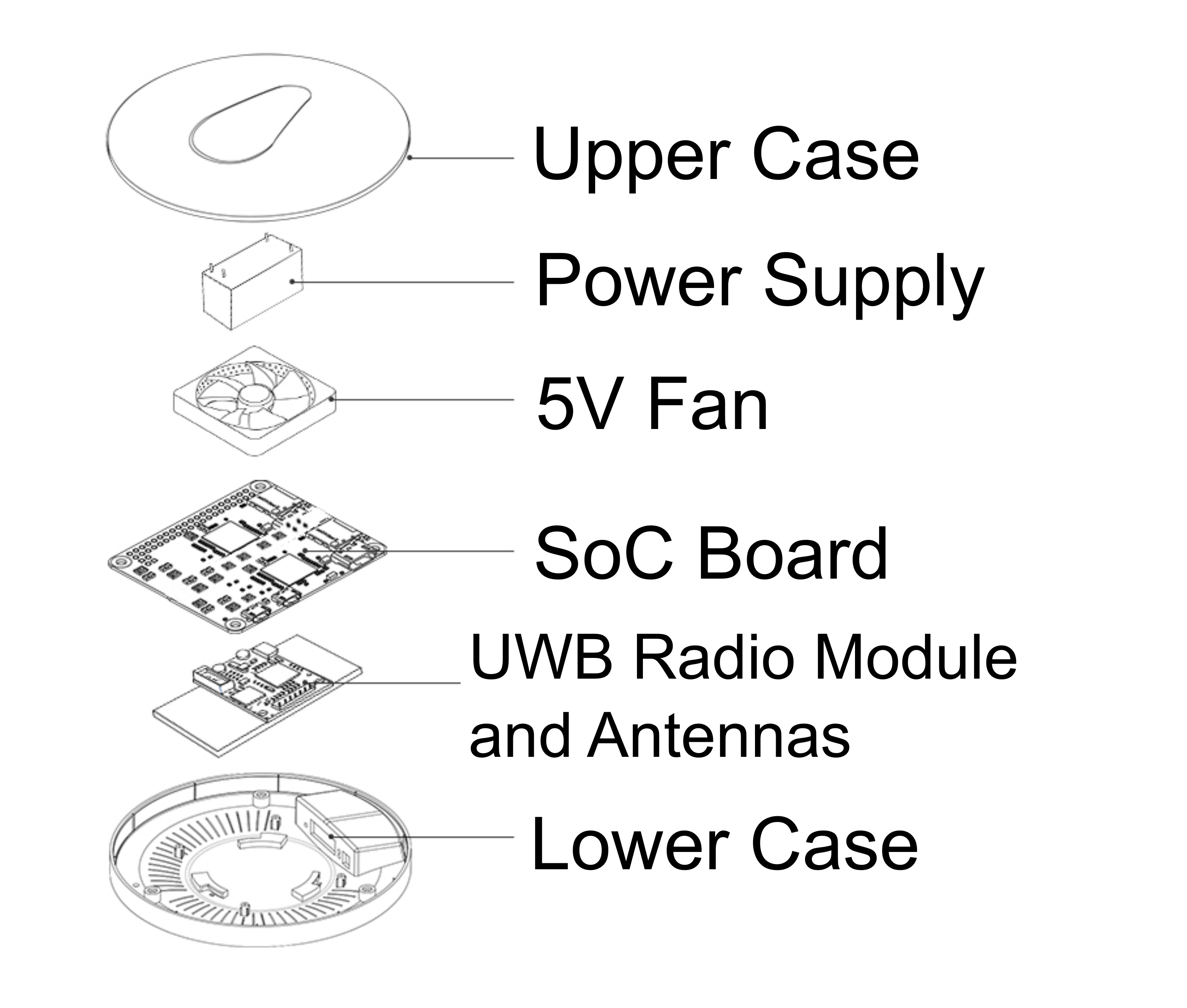}
		\caption{\systemname's hardware components.}
	\label{fig:explode}
	\end{minipage}
	\hspace{2ex}
	\begin{minipage}[t]{0.33\linewidth}
		\centering
		\includegraphics[width=.96\linewidth]{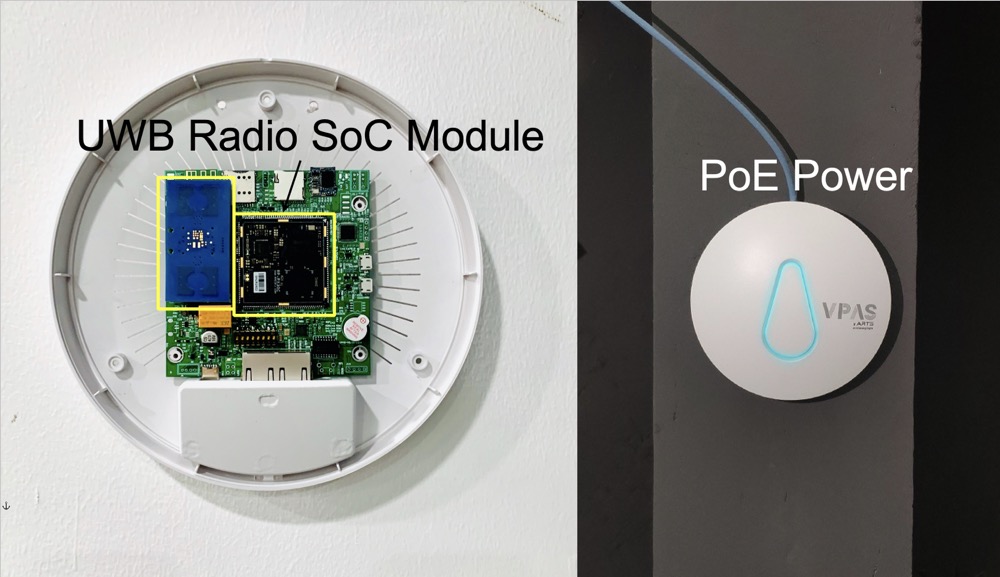}
		\caption{The PCBs of UWB radio and SoC board.}
	\label{fig:real_sys}
	\end{minipage}
	\hspace{6ex}
	\begin{minipage}[t]{0.23\linewidth}
		\centering
		\includegraphics[width=.63\linewidth]{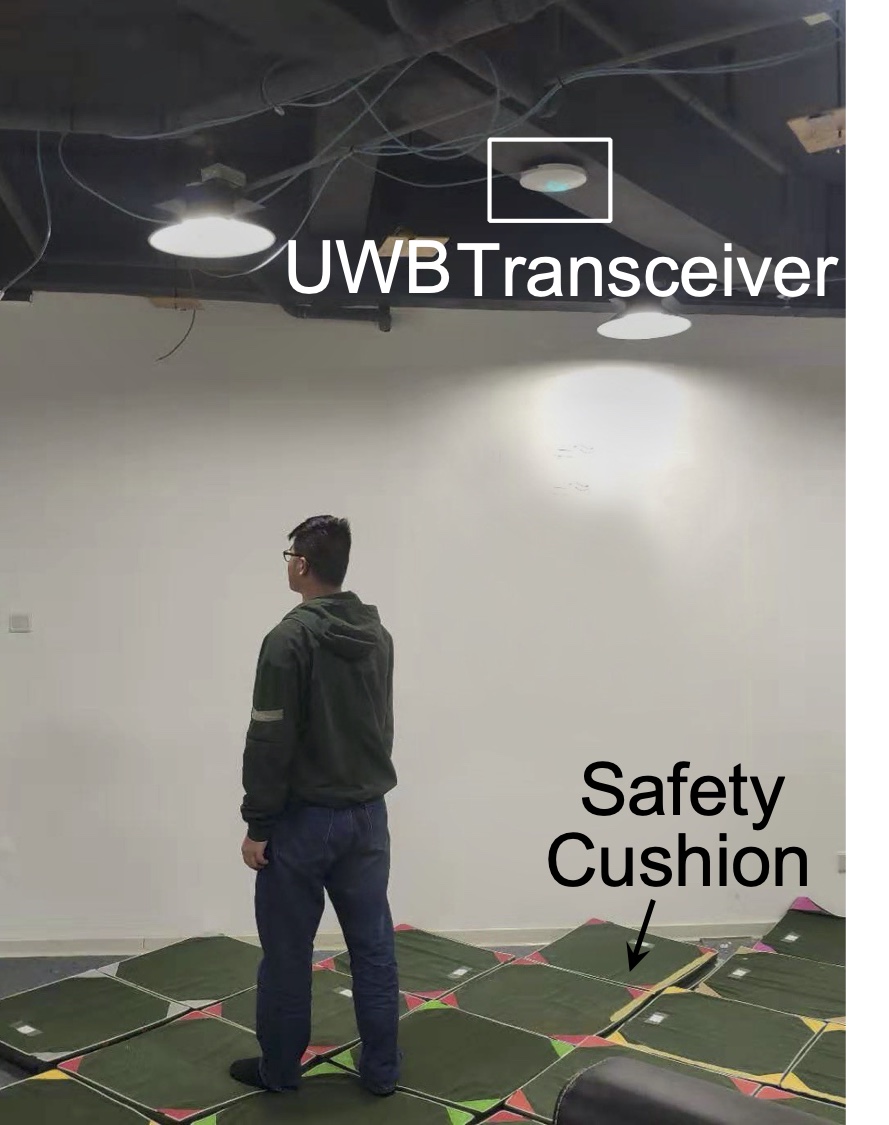}
		\caption{Experiment setup.}
		\label{fig:eval_setup}
	\end{minipage}
\end{figure*}

Our hardware prototype includes a power supply, a 5V fan, an SoC~(System on Chip) module with Rockchip 3308~\cite{rockpi} and a UWB radio module as shown in Fig.~\ref{fig:explode}.  We employ a cheap commodity UWB radio XETHRU~\cite{b17} to transmit and receive UWB signals. The UWB radio is connected to the edge device~(SoC board) via Serial Peripheral Interface~(SPI). The hardware PCB is small with a size of $10.1 \times 10.6 \ cm^2$ illustrated in Fig.~\ref{fig:real_sys}. 
Our signal adapted CNN model is implemented on TensorFlow~\cite{b45}, and our model is converted into a compressed flat buffer with 32-bit floats using TensorFlow Lite. Thus, the model can be deployed on mobile and edge devices. Note that for fair comparisons with other models, we also convert other models via TensorFlow Lite. %

\subsection{Evaluation} \label{sec:eval}

\begin{figure*}[t]
			\centering
		\includegraphics[width=\linewidth]{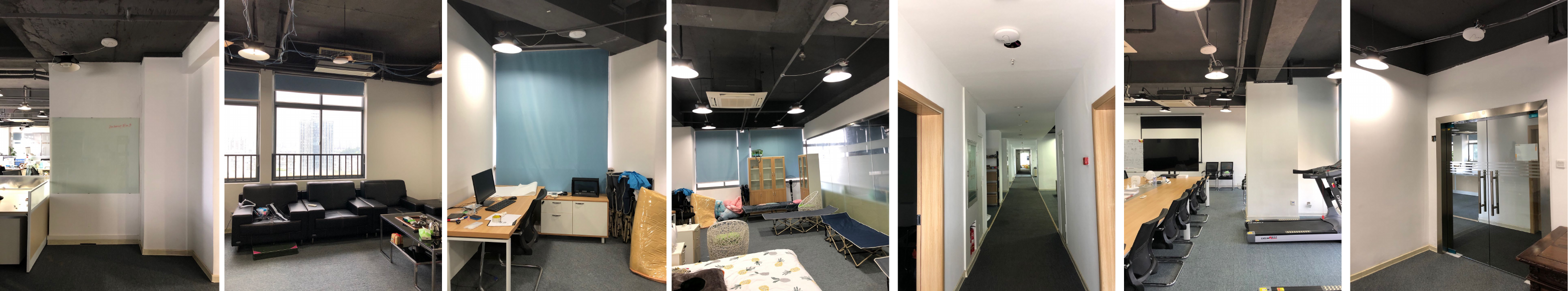}
		\caption{Seven environments where training and testing datasets are collected.}
		\label{fig:envs}
		\vspace{-1ex}
\end{figure*}
\subsubsection{Evaluation Setup}
To test the generalization of our system, we ensure that the training data and test data are different except for the comparison study.  It means that we collect data from 7 environments shown in Fig.~\ref{fig:envs}, and only data from two environments are used for training. The data of the rest 5 environments are used for testing. We collected a large data set. The number of training samples~(activities) is 15,000, and that of testing samples~(activities) is 40,000. 
Seven commonly seen activities are considered in this paper including bending (B), falling (F), lying down (L), standing up (SU), sitting down (SD), squatting down (SQ), and walking (W).  
 The \systemname\ hardware is mounted on the ceiling to classify the activities as shown in Fig.~\ref{fig:eval_setup}. The height of mounted UWB transceiver is about 2.7$\!~$m above the ground.

To evaluate our system, we use \textit{Accuracy}, \textit{Precision}, \textit{Recall} and \textit{F1 score} as the metrics. For simplicity, we use TP, FN, TN, and TP to represent True Positives, False Negatives, True Negatives and False Positives respectively. \textit{Precision} is the ratio of the number of correctly classified activities to the number of all classified activities i.e., $p = \frac{TP}{TP+FP}$. \textit{Recall} is the fraction of correctly classified activities over all activities of that, i.e., $ r = \frac{TP}{TP+FN} $. 
\textit{F1 score}~\!$=\frac{2pr}{p+r} $ is the harmonic mean of precision and recall. 

\subsubsection{Activity Recognition}
We evaluate the performance of \systemname\ with new targets in new environments which are not included in the training process. In this experiment, we use a $3 \times 3$ kernel, and to emulate the the real scenario, we mix a large amount of no human activity
samples with activity samples. We plot the results in TABLE~\ref{tab:sys_eval}. Even \systemname\ does not train in the new environments with new human targets, the results show that \systemname\ can still achieve an average of 0.965 in \textit{recall} and 0.969 in \textit{precision}. %
\systemname\ demonstrates the capability of working with new targets in new environments without further training.  %
We also compare our design with other state-of-the-art schemes such as XGBoost~\cite{b54} and SVM that are not based on neural network. We train all models with the same training sets. 
We can clearly see from TABLE~\ref{tab:sys_eval} that \systemname\ achieves much better performance than XGBoost and SVM in terms of all the metrics. Although XGBoost also performs reasonably well, \systemname\ outperforms it because CNN-based \systemname\ can capture complex time-frequency patterns in high-dimensional data input.
\begin{table}[ht] 
	\centering	
	\caption{\systemname's average evaluation results in UWB radio.}
	\begin{tabular} {c c c c c}
		\hline
		\space  & Precision & Recall & F1 Score  \\
		\hline
		\systemname &  \textbf{0.969} & \textbf{0.965} & \textbf{0.967}  \\
		XGBoost \cite{b54} &  {0.851} & {0.856} & {0.852}  \\
		SVM~(Linear) &  {0.452} & {0.460} & {0.455}  \\
		\hline
		\label{tab:sys_eval}
	\end{tabular}
\end{table}

\begin{figure} [t]
	\setlength\abovecaptionskip{-1pt}
	\setlength\belowcaptionskip{-1pt}
	\begin{minipage}[t]{0.48\linewidth}
		\centering
		\includegraphics[width=1\textwidth]{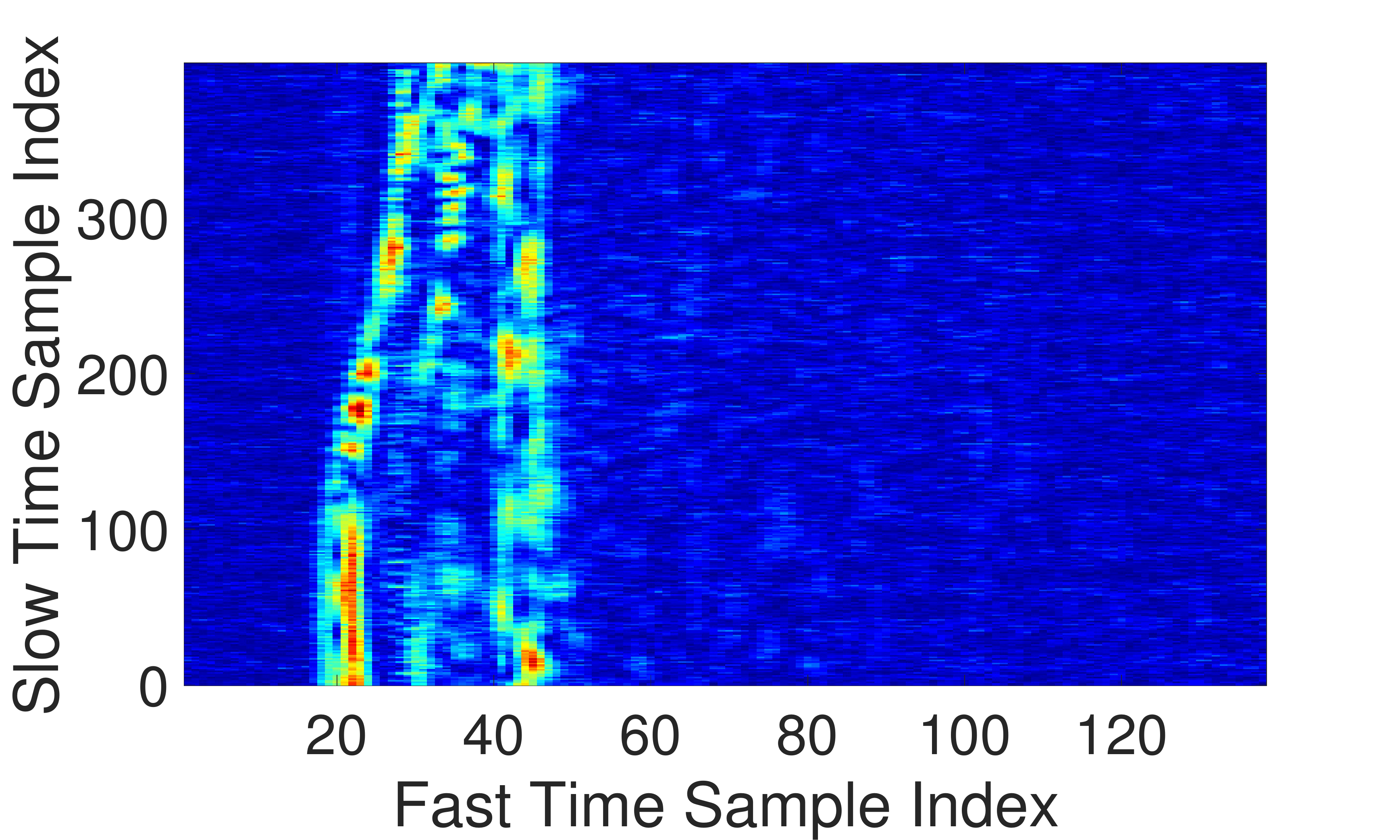}
		\caption{The power delay profile of sitting down.}
		\label{fig:sit_time}
		\hspace{-0.05cm}
	\end{minipage}
	\begin{minipage}[t]{0.48\linewidth}
		\centering
		\includegraphics[width=1\textwidth]{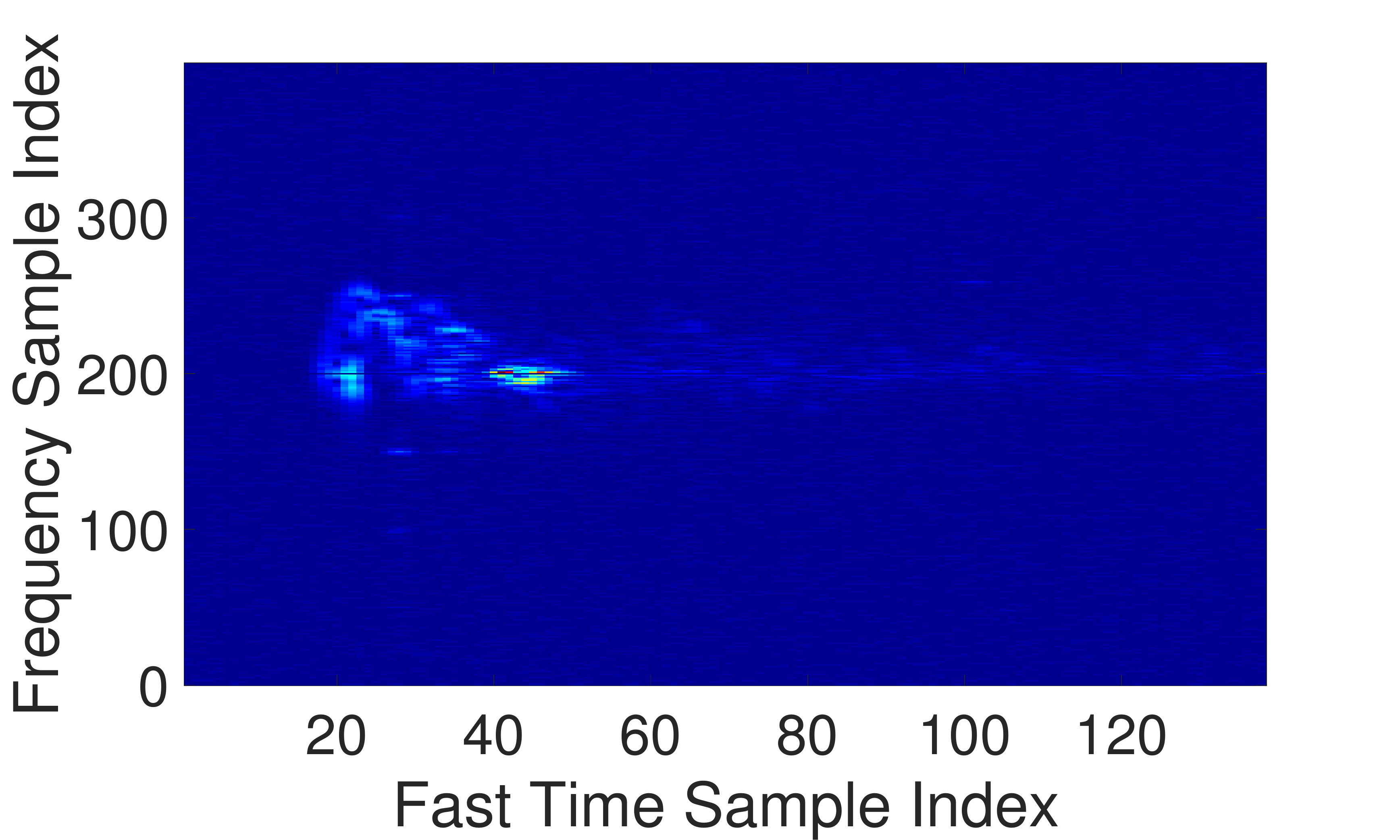}
		\caption{The Doppler-Range profile of sitting down.}
		\label{fig:sit_fft}
	\end{minipage}
	\vspace{-2ex}
\end{figure}

\begin{figure} [t]
	\setlength\abovecaptionskip{-1pt}
	\setlength\belowcaptionskip{-1pt}
	\begin{minipage}[t]{0.48\linewidth}
		\centering
		\includegraphics[width=1\textwidth]{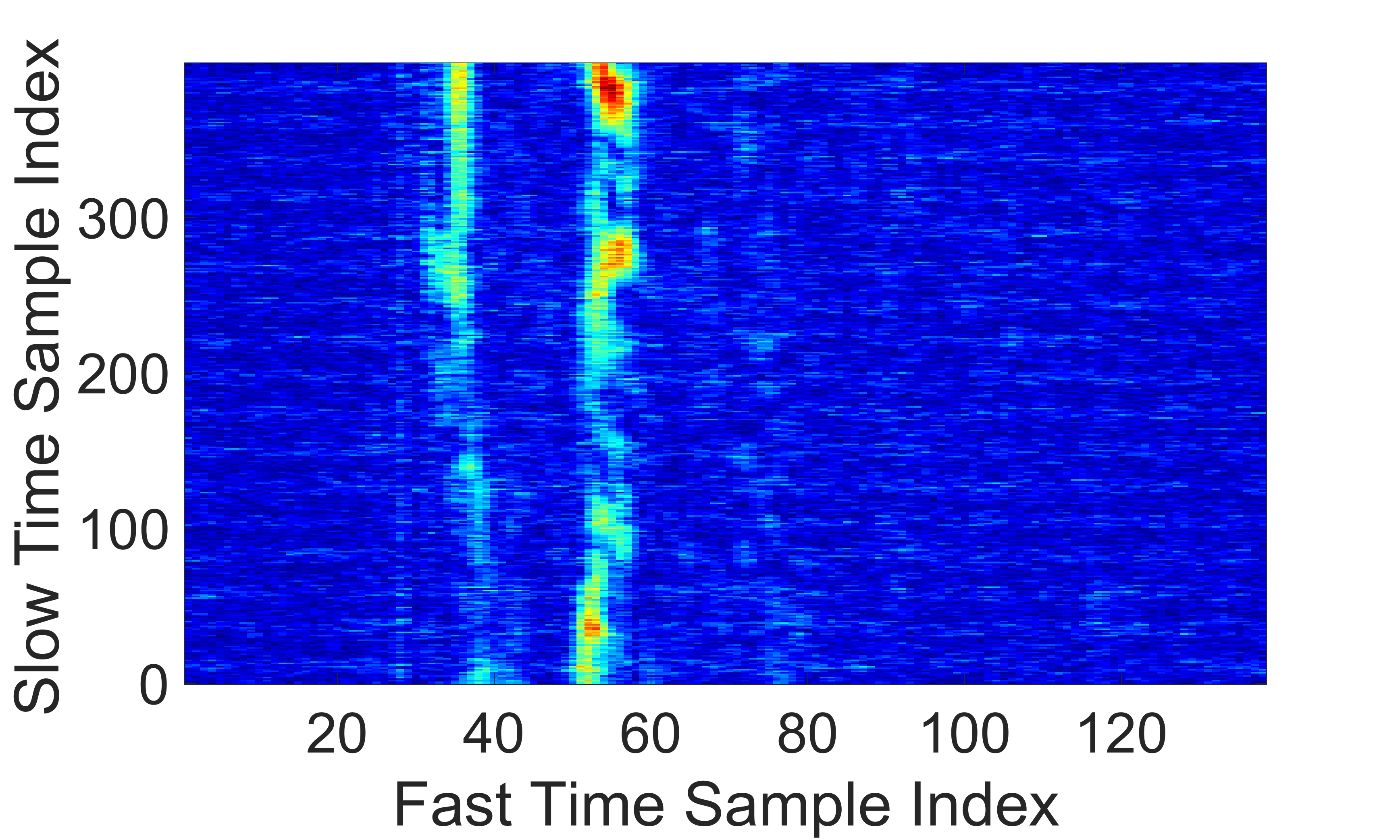}
		\caption{The power delay profile of squatting down.}
		\label{fig:squat_time}
		\hspace{-0.05cm}
	\end{minipage}
	\begin{minipage}[t]{0.48\linewidth}
		\centering
		\includegraphics[width=1\textwidth]{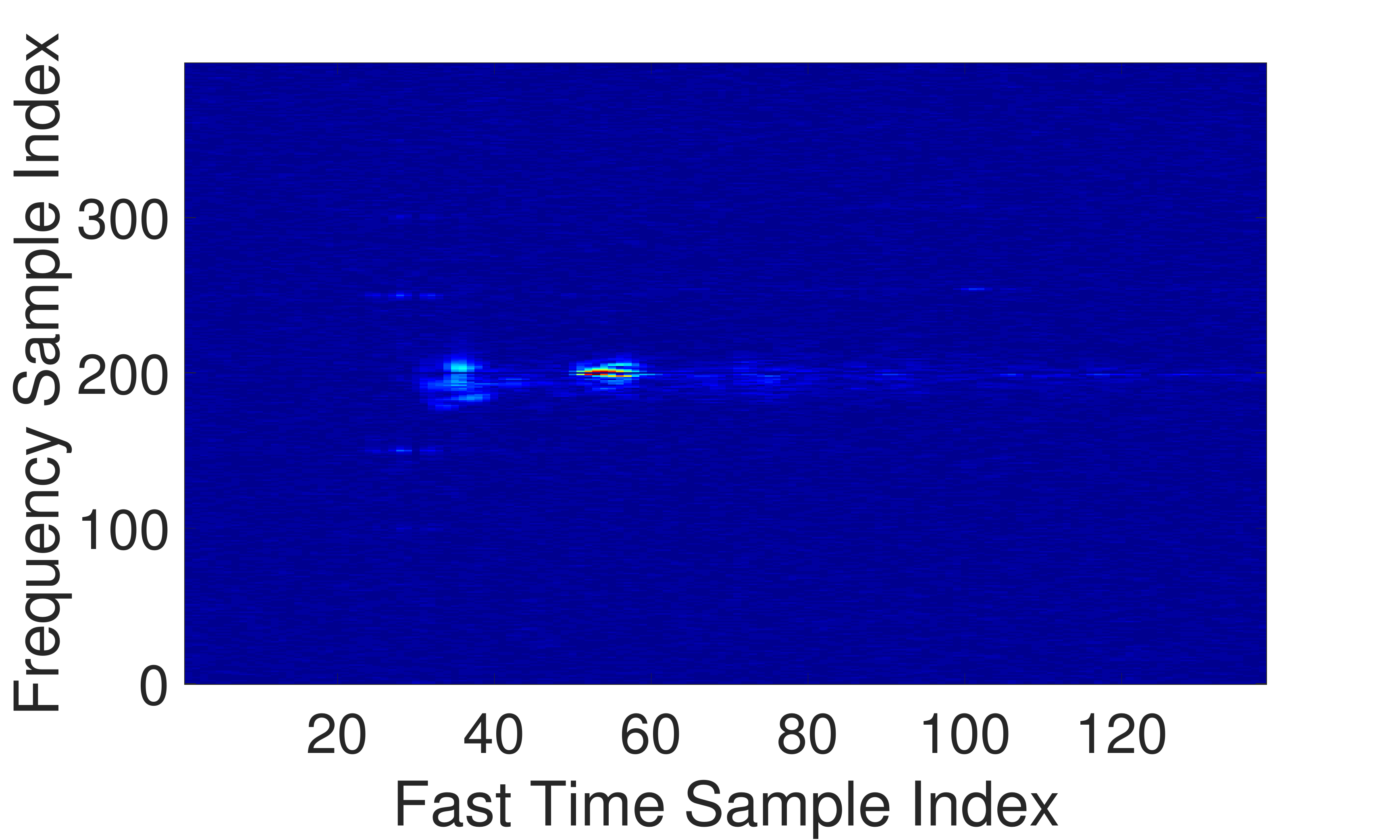}
		\caption{The Doppler-Range profile of squatting down.}
		\label{fig:squat_fft}
	\end{minipage}
\end{figure}

\begin{table}[t] 
\vspace{-1ex}
	\centering	
	\caption{\systemname's average evaluation results in Wi-Fi.}
	\begin{tabular} {c c c c c}
		\hline
		\space  & Precision & Recall & F1 Score  \\
		\hline
		\systemname~(Wi-Fi) &  \textbf{0.792} & \textbf{0.800} & \textbf{0.796}  \\
		CrossSense (Wi-Fi)  \cite{b47}   & 0.671& 0.618 & 0.643 \\
		\hline
		\label{tab:sys_eval_wifi}
	\end{tabular}
	\vspace{-3ex}
\end{table}
We further compare \systemname\ with the state-of-the-art Wi-Fi-based system. We implement Wi-Fi-based CrossSense~\cite{b47} in which
STFT-like  analysis is used to extract features. 
We mount one Wi-Fi transmitter equipped with Intel 5300 card on the ceiling to transmit, and employ another Wi-Fi device equipped with Intel 5300 card to receive signals on the floor. The Wi-Fi transmitter is mounted 2.7$\!~$m above the ground.
Note that CrossSense employed multiple transmitters and receivers while we use only one transmitter and receiver. %
While CrossSense achieves a good performance with their own dataset which employs multiple transceiver pairs, with our dataset with only one transceiver pair, we can see that \systemname\ achieves a much better performance than the state-of-the-art CrossSense as shown in TABLE~\ref{tab:sys_eval_wifi}. The reason is that even though CrossSense employs transfer learning to achieve cross-site sensing capability, the unique advantage of \systemname\ is the motion detection module. The data samples without any movement can be easily detected and removed via motion detection module, and \systemname\ only focuses on those data samples with human activities. On the other hand, CrossSense is not able to exclude those non-activity samples. More importantly, \systemname\ with UWB radio is able to provide a much higher time delay resolution and thus finer-grained Doppler information can be obtained. Due to the 40~\!MHz narrow-band, the time delay resolution of of Wi-Fi Intel 5300 card is 25~\!ns, and the corresponding distance resolution is as large as 7.5~\!m. It is thus very hard to distinguish the miniscule motions of body parts with similar distance with respect to the sensing hardware. Moreover, when the Wi-Fi card is mounted on the ceiling, the Doppler shifts caused by velocities of human sitting down and squatting down are very similar. We plot the time and frequency spectrograms of the UWB signals for squatting down and sitting down from Fig. \ref{fig:sit_time} to \ref{fig:squat_fft}. We can see that the spectrograms of both time and frequency can be employed to easily distinguish sitting down and squatting down. 
\textcolor{black}{It demonstrates that HAR-SAnet can extract rich features to enable a fine-grained HAR.}


\subsubsection{Impacts of Kernel Sizes}
We also evaluate the effect of different kernel sizes on \systemname. The results are shown in TABLE~\ref{tab:kernel_size}. The interesting insight here is when the kernel size increases, the performance improves first. However, if we keep increasing the kernel size to $9 \times 9$, the precision starts decreasing. 
The reason is that a larger kernel size has a larger receptive field, hence \systemname\ with a larger kernel size can capture more features.  But if the receptive field is too large, \systemname\ will end up capturing useless noise in the spectrograms, thus the performance degrades.
\begin{table}[h] 
	\centering	
	\caption{Different convolution kernel sizes impact on \systemname.}
	\begin{tabular} {c c c c c}
		\hline
		\space Kernel size & Precision & Recall & F1 Score  \\
		\hline
		$3\times3$ &  {0.969} & {0.965} & {0.965}  \\
		$5\times5$&  {0.980} & {0.980} & {0.978}  \\
		$7\times 7 $&  {0.984} & {0.980} & {0.984}  \\
		$9\times 9 $&  {0.982} & {0.979} & {0.982}  \\
		\hline
		\label{tab:kernel_size}
	\end{tabular}
\end{table}

\subsubsection{Impacts of Height}
We mount UWB transceiver at different heights including 2.2$\!~$m, 2.7$\!~$m, and 3.5$\!~$m to evaluate the system performance. We train the model with a $3 \times 3$ kernel size using the datasets collected at the height of 2.7$\!~$m, and test the performance at the other two heights. 
\begin{table}[ht] 
	\centering	
	\caption{Different heights impact on \systemname.}
	\begin{tabular} {c c c c c}
		\hline
		\space  Height & Precision & Recall & F1 Score  \\
		\hline
		2.2$\!~$m&  {0.957} & {0.957} & {0.957}  \\
		2.7$\!~$m &  {0.969} & {0.965} & {0.965}  \\
		3.5$\!~$m&  {0.930} & {0.922} & {0.926} \\
		\hline
		\label{tab:height}
	\end{tabular}
\end{table}
The results are shown in TABLE~\ref{tab:height}. All results for height 2.2$\!~$m are very close to those for 2.7$\!~$m. However, for the height of 3.5$\!~$m, the results are slightly worse. We believe the reason is that the reflected RF signals from human body become weaker when the sensing device is mounter higher,  %
introducing slightly more errors in \systemname.

\subsubsection{Motion Detection}
We use two metrics to evaluate the accuracy of motion detection: True Positive Rate (TPR) and False Alarm Rate (FAR). TPR is the ratio between the number of times when \systemname\ correctly detects the human motion and the total number of observed motions. FAR is the ratio of the number of times when \systemname\ wrongly detects a motion to the number of times when there is no motion. 
\begin{figure}[h]
	\centering
	\includegraphics[width=0.65\linewidth]{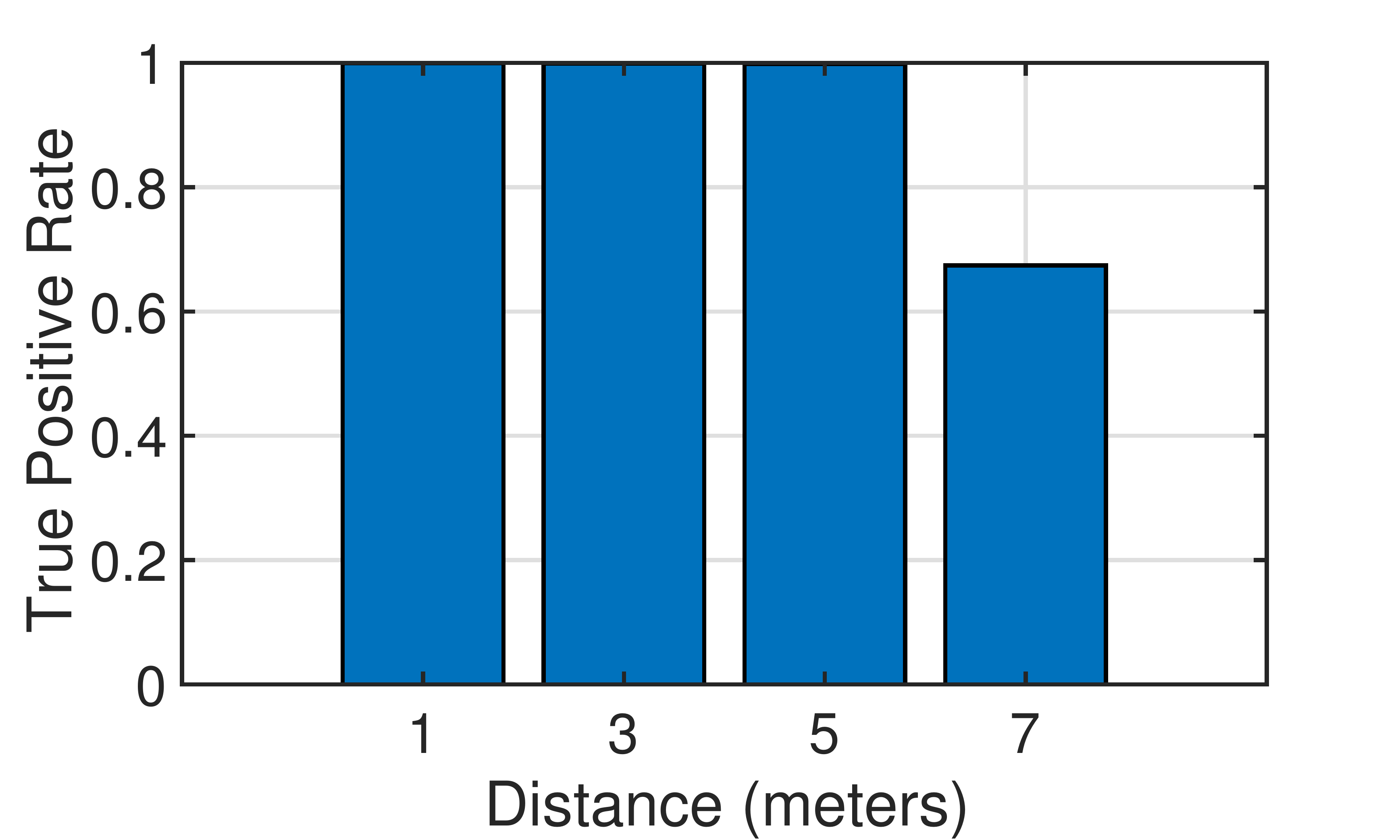}
	\caption{Motion detection range with true positive rate.}
	\label{fig:tpr}
\end{figure}

\systemname\ can detect motions successfully within the horizontal range of 5 meters.  When a target enters this range, his or her motion can be accurately detected. The TPR results are shown in Fig.~\ref{fig:tpr}. We calculate each TPR bar in Fig.~\ref{fig:tpr} using 200 activity samples collected at different positions but with the same distance. The target moves away from the sensing hardware from 1m. We can see that as long as the distance between the target and the sensing device is below 5m, the motion detection accuracy is always 100\%. We the distance is increased to 7 meters, the TPR value starts to decrease. The reason is that the transmission power of all UWB devices is regulated by the Federal Communications (FCC) in the US and the European Telecommunications Standards Institute (ETSI) in the Europe. The maximum allowed mean equivalent isotropically radiated power (EIRP) spectra density is -41.3 dBm/MHz~\cite{b35}, and is only around 0.1\% of the density allowed for Wi-Fi~\cite{b36,b37}. Thus, the radio coverage range is now in the scale of room level. In the future, we plan to explore the possibility of employing LoRa signal to significantly increase the sensing range to building level.  

We also measure the FAR of  HAR-SAnet. The FAR is very low at a rate of 0.083 false alarms per hour. We record the human motions to measure the FAR in 24 hours a day when no one moves in the sensing range. During one week, there are only a total of 14 false alarms.  We believe these rare false alarms are due to the suddenly increasing noise levels. 

\subsubsection{Ablation Study}
Our signal adapted CNN structure includes two branches that deal with time and frequency domain features, respectively. To better understand the effectiveness of our model, we need to conduct ablation study under the same experiment condition. 
\begin{figure*}[t]
	\setlength\abovecaptionskip{-1pt}
	\setlength\belowcaptionskip{-1pt}
	\begin{minipage}[t]{0.33\linewidth}
		\centering
		\includegraphics[width=\linewidth]{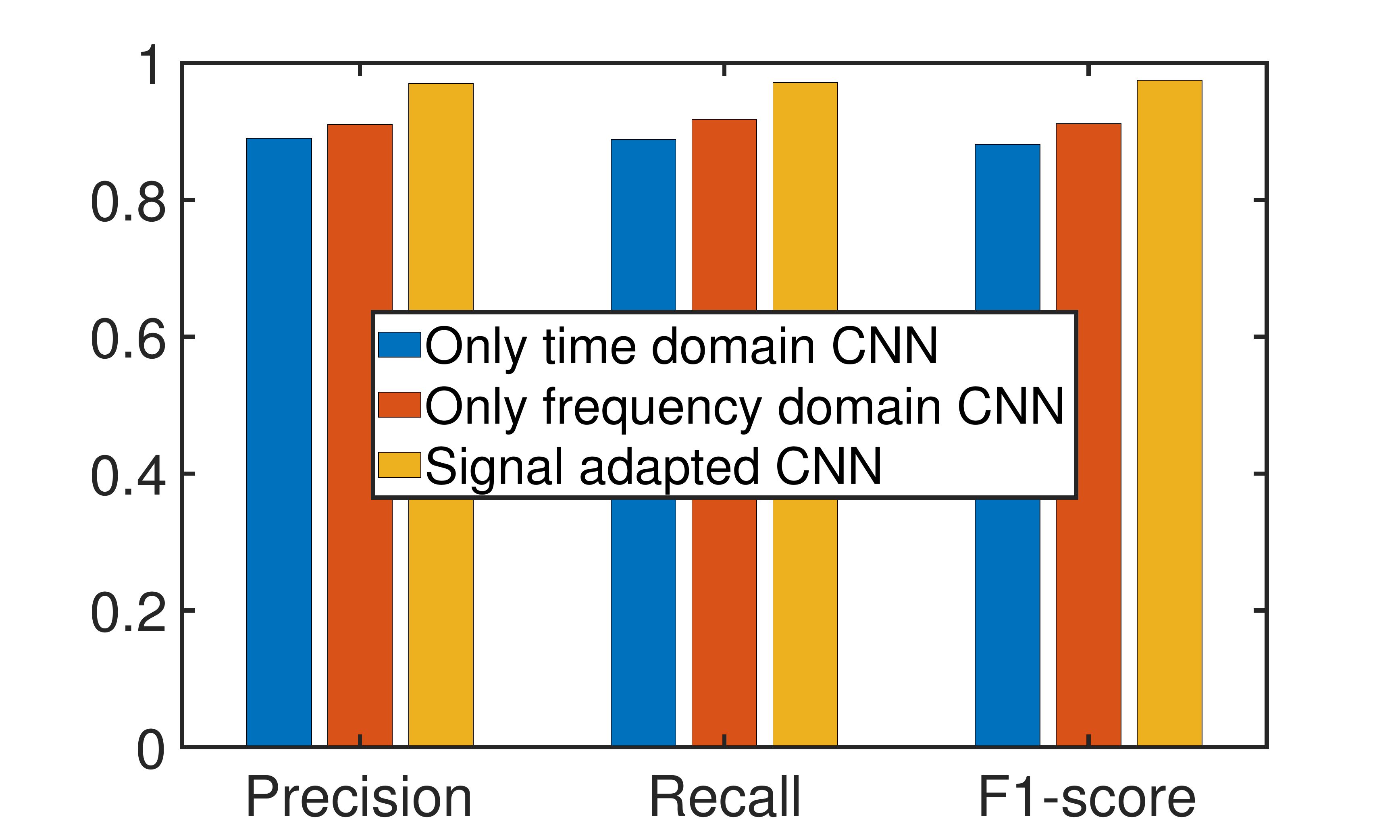}
		\caption{Ablation study on each module of signal  adapted convolutional neural network.}
		\label{fig:ablation}
	\end{minipage}
	\begin{minipage}[t]{0.33\linewidth}
		\centering
		\includegraphics[width=\linewidth]{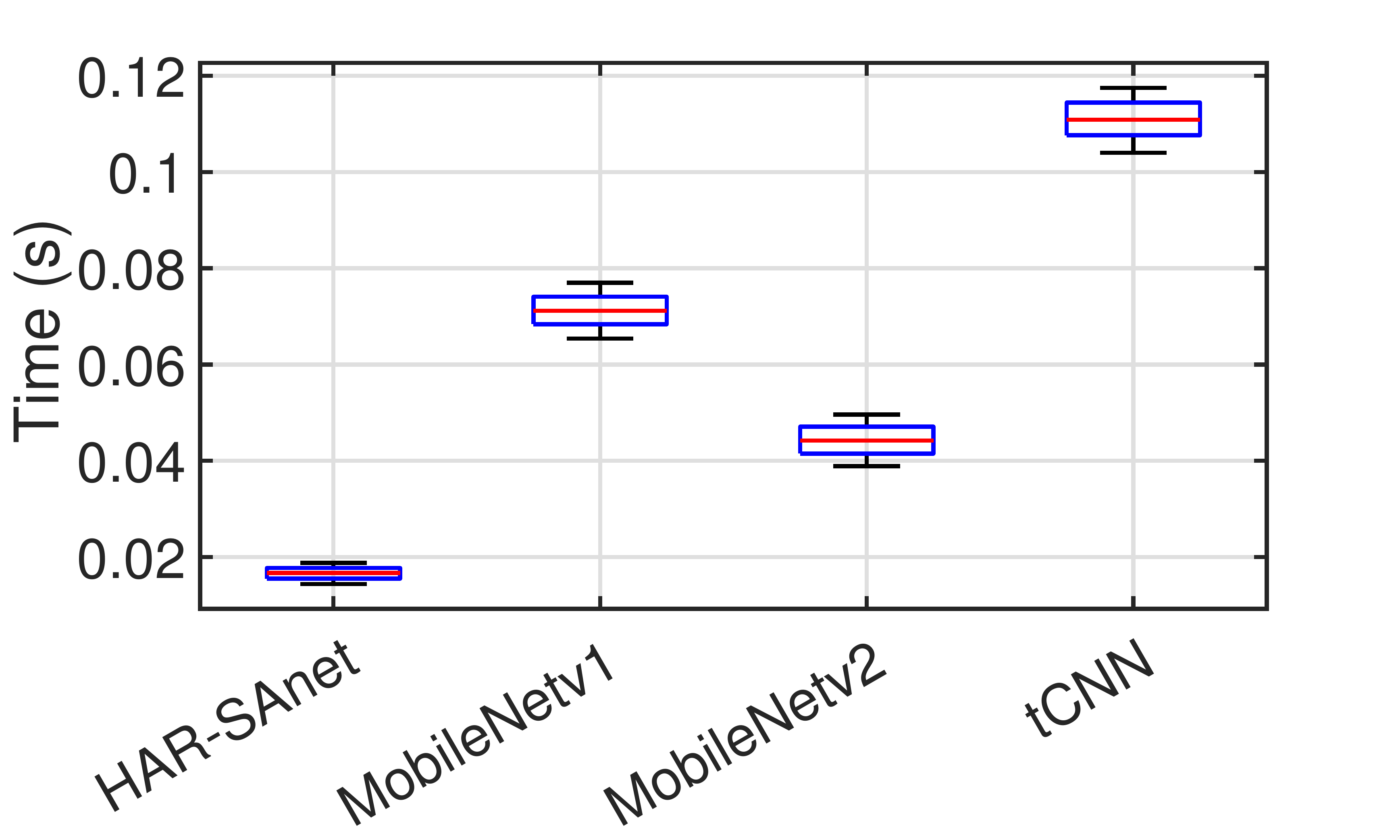}
		\caption{Execution time in edge device.}
		\label{fig:tflite_speed}
	\end{minipage}
	\begin{minipage}[t]{0.33\linewidth}
		\centering
		\includegraphics[width=\linewidth]{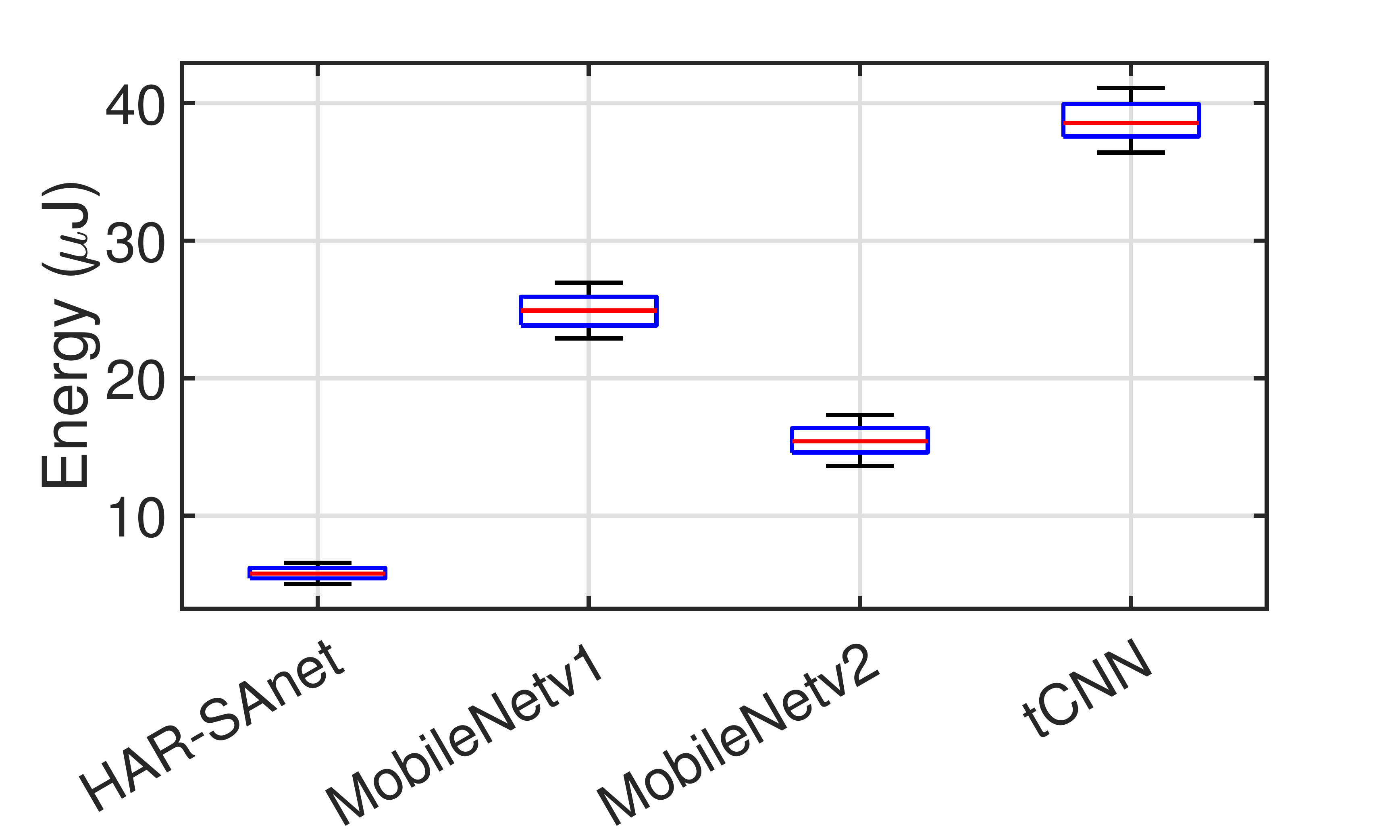}
		\caption{Average energy cost per inference.}
		\label{fig:energy}
	\end{minipage}
\end{figure*}
To fairly study each branch of our model, we use the same signal adapted CNN architecture to evaluate a single domain. The results are illustrated in Fig. \ref{fig:ablation}. The \textit{precision} of time domain is 0.88, and that of the frequency domain is 0.91.  Our signal adapted model combines both time and frequency domains into two branches, and is able to obtain a much higher \textit{precision} of 0.97. Similar to \textit{precision}, the \textit{recall} and \textit{F1 score} of our signal adapted model outperform other single domain designs. Our model improves over single time and single frequency domain by an average of 10\% and 6.5\%, respectively. Hence, considering information from both time and frequency domains is effective in improving the performance of the neural network design.

\subsubsection{ Comparison with the state-of-the-arts}
\textcolor{black}{In this section, we study the efficiency of various CNN blocks in \systemname.
We compare the CNN block adopted by \systemname\ with three state-of-the-art baseline blocks: MobileNetv1~\cite{b41}, MobileNetv2~\cite{b46} and a traditional CNN (tCNN) with three layers~\cite{b53}. 
We train all these models with the same datasets, and test them on edge device in real time scenarios.} MobileNetv1 and MobileNetv2 are state-of-the-art CNN design for mobile and edge devices from Google. 
We use them to replace the CNN block in \systemname, and measure the execution time and energy cost of the whole system. 
We use TensorFlow Lite to compress \systemname, MobileNetv2, MobileNetv1, and tCNN into 8-bits representations.  We use SoC module to infer 1000 activities and record the execution time of each inference. The boxplots of those model execution time  and energy cost are shown in Fig.~\ref{fig:tflite_speed}, and Fig.~\ref{fig:energy}, respectively. We also present the accuracy, average execution time, and energy cost per inference in TABLE~\ref{tab:acc_complex}, where \textit{accuracy} is the ratio of \textcolor{black}{correctly classified activity samples} over all samples. It is observable that \systemname\ achieves a comparable (actually slightly better) accuracy while bearing a much lower complexity (8 to 3 times lower than others).
\begin{table}[h] 
	\centering	
	\caption{Comparison of several network architectures over complexity (time and energy) and classification accuracy.}
	\begin{tabular} {c c c c }
		\hline
		\space  & Complexity (s) & Energy ($\mu$J) & Accuracy\\
		\hline
        \systemname &  \textbf{0.016} & \textbf{5.80}  &\textbf{0.974}  \\
		MobileNetv1 \cite{b41} & 0.071 & 15.40 & 0.960  \\
		MobileNetv2 \cite{b46} & 0.044 & 24.92 & 0.963  \\
		Traditional CNN  & 0.111 &  38.57    &  0.967 \\
		\hline
		\label{tab:acc_complex}
	\end{tabular}
\end{table}

\textcolor{black}{We have also looked at the theoretical complexity characterized by the Float-Point OPerations (FLOPs) metric, which has surprisingly shown a similar count (around 0.08 million FLOPs) for all four CNN blocks. It appears that, though different CNN blocks share similar FLOPs, their runtime complexity differ a lot.} 
The reason is that the CNN computing is not only determined by the computing operations, but also by memory swap. For instance, tCNN spends more time in memory swap than the other blocks designed specifically for running on edge device.

\textcolor{blue}{
}

\section{Related Work} \label{sec:rw}
Past work on activity recognition can be grouped into two categories: wearable-based and non-wearable-based schemes.  For wearable-based schemes, notable examples include smartphones and accelerometers \cite{b21,b22,b23}.  However, people, especially the elderly are usually reluctant to wear wearables because of skin irritation and they often forget to wear the devices~\cite{b24, b25}. On the other hand, non-wearable scheme was proposed to address the above limitations. %
Camera-based solutions \cite{b26,b27} can achieve accurate activity recognition, but the privacy and narrow field of view are the issues hindering their wide deployment. Audio-based solutions~\cite{b28,b58,b59} can achieve highly accurate sensing performance due to the low propagation speed in the air. However, these systems are vulnerable to the acoustic noise and interference around us and the sensing range is usually very limited~(below 1m).

Our work is most related to RF-based solutions. Existing work on device-free HAR can be divided into three categories: Received Signal Strength Indicator (RSSI)-based, CSI-based and radar-based solutions. 
The RSSI-based solutions rely on the fact that the human activities can cause signal strength change. RSSI-based HAR systems leverage the unique signal strength changes to classify activities \cite{b2,b29,b30}. However, since the RSSI readings are very coarse,  such systems can only recognize the coarse-grained human activities, and the achieved accuracy is relatively low. %

 Recently, CSI-based solutions  have attracted a lot of attentions in RF-based HAR \cite{b2,b3,b4,b5,b6,b7,b8,b9,ziqi_fall_2018}. These solutions apply the STFT or wavelet transforms to estimate the signal changes caused by target velocity~\cite{b3,b6}. They expect that the velocities of different body parts can be used to classify the activities via machine learning or deep learning. However, CSI-based solutions pose some significant limitations. For instance, the limited Wi-Fi bandwidth cannot separate reflections from different body parts. Hence, the features which can be used to distinguish activities are limited. Moreover, CSI readings can only be retrieved from two types of commodity 802.11n Wi-Fi cards (Intel 5300 \cite{b31} and Atheros 9390 network interface cards (NICs) \cite{b32}). %

 Radar technology is also leveraged to classify human activities \cite{b11,b12,b13,b14,b16,b20,zhe_rfnet_2020}. The authors in~\cite{b11,b12,b13,b14} use one-dimensional feature such as Micro-Doppler or Doppler information to recognize human activities. Thus, same as CSI-based solutions, one-dimensional feature limits the performance of HAR. Some other work  \cite{b16,b20,b33} employ a specialized hardware, USRP to implement a Frequency-Modulated Continuous-Wave (FMCW) radar system with a large antenna array to classify human activities and demonstrate high accuracy of activity recognition. However, these specialized devices are usually expensive and there is a huge gap in terms of price and functionality between the software-defined radio hardware platform and cheap COTS hardware. We would like to realize HAR with cheap commodity hardware.  %

Furthermore, most of existing research work employ a powerful computer to realize HAR.  It is not always practical because most edge devices have a limited storage and a limited computational power. \systemname\ is not only a COTS solution but also designs lightweight neural network model for resource-constrained edge devices. Last but not least, we believe that the UWB radio can be further utilized to drive other sensing applications, such as replacing Wi-Fi for indoor localization~\cite{m3,luo2014ilocscan} and already being applied to vibration and vital sign monitoring~\cite{zheng2020v2ifi, UWHear}.

\section{Conclusion} \label{sec:conclusion}
In this paper, we propose a HAR system hosted on COTS UWB radio. Owing to the large bandwidth of UWB radio, our system can obtain richer motion features from RF signals compared to Wi-Fi-based solutions. To make our system work with resource-constrained edge device, a signal adapted convolutional neural network model is designed to extract features and classify activities without handcraft. The system is evaluated in multiple real-life environments and comprehensive experiments  demonstrate that \systemname\ can obtain a precision of 96.9\%  and a recall of 96.5\%. %
We believe the proposed methods can benefit a large range of other sensing applications. In the future, we plan to extend \systemname\ to MIMO UWB radio systems to explore the boundary of the sensing capability. %

\ifCLASSOPTIONcaptionsoff
  \newpage
\fi

\bibliographystyle{IEEEtran}
\bibliography{tmc}

\begin{thebibliography}{10}
\providecommand{\url}[1]{#1}
\csname url@samestyle\endcsname
\providecommand{\newblock}{\relax}
\providecommand{\bibinfo}[2]{#2}
\providecommand{\BIBentrySTDinterwordspacing}{\spaceskip=0pt\relax}
\providecommand{\BIBentryALTinterwordstretchfactor}{4}
\providecommand{\BIBentryALTinterwordspacing}{\spaceskip=\fontdimen2\font plus
\BIBentryALTinterwordstretchfactor\fontdimen3\font minus
  \fontdimen4\font\relax}
\providecommand{\BIBforeignlanguage}[2]{{%
\expandafter\ifx\csname l@#1\endcsname\relax
\typeout{** WARNING: IEEEtran.bst: No hyphenation pattern has been}%
\typeout{** loaded for the language `#1'. Using the pattern for}%
\typeout{** the default language instead.}%
\else
\language=\csname l@#1\endcsname
\fi
#2}}
\providecommand{\BIBdecl}{\relax}
\BIBdecl

\bibitem{b24}
R.~Steele, A.~Lo, C.~Secombe, and Y.~K. Wong, ``{Elderly Persons’ Perception
  and Acceptance of using Wireless Sensor Networks to Assist Healthcare},''
  \emph{International Journal of Medical Informatics}, pp. 788 -- 801, 2009.

\bibitem{b48}
G.~Ogbuabor and R.~La, ``{Human Activity Recognition for Healthcare Using
  Smartphones},'' in \emph{Proc. of the 10th IEEE ICMLC}, 2018, p. 41–46.

\bibitem{b20}
Y.~Tian, G.-H. Lee, H.~He, C.-Y. Hsu, and D.~Katabi, ``{RF-Based Fall
  Monitoring Using Convolutional Neural Networks},'' \emph{Proc. of the 17th
  ACM UbiComp}, vol.~2, 2018.

\bibitem{b25}
F.~{Sposaro} and G.~{Tyson}, ``{iFall: An Android Application for Fall
  Monitoring and Response},'' in \emph{Proc. of the IEEE Engineering in
  Medicine and Biology Society}, 2009, pp. 6119--6122.

\bibitem{b28}
Y.~{Li}, K.~C. {Ho}, and M.~{Popescu}, ``{A Microphone Array System for
  Automatic Fall Detection},'' \emph{IEEE Transactions on Biomedical
  Engineering}, vol.~59, pp. 1291--1301, 2012.

\bibitem{b1}
T.~W. Hnat, V.~Srinivasan, J.~Lu, T.~I. Sookoor, R.~Dawson, J.~Stankovic, and
  K.~Whitehouse, ``{The Hitchhiker's Guide to Successful Residential Sensing
  Deployments},'' in \emph{Proceedings of the 9th ACM SenSys}, 2011, p.
  232–245.

\bibitem{b49}
A.~{Jalal}, M.~Z. {Uddin}, and T.~. {Kim}, ``{Depth video-based Human Activity
  Recognition System using Translation and Scaling Invariant Features for Life
  Logging at Smart Home},'' \emph{IEEE Transactions on Consumer Electronics},
  pp. 863--871, 2012.

\bibitem{b2}
S.~{Sigg}, M.~{Scholz}, S.~{Shi}, Y.~{Ji}, and M.~{Beigl}, ``{RF-Sensing of
  Activities from Non-Cooperative Subjects in Device-Free Recognition Systems
  Using Ambient and Local Signals},'' \emph{IEEE Transactions on Mobile
  Computing}, vol.~13, no.~4, pp. 907--920, 2014.

\bibitem{b3}
W.~Wang, A.~X. Liu, M.~Shahzad, K.~Ling, and S.~Lu, ``{Understanding and
  Modeling of WiFi Signal Based Human Activity Recognition},'' in \emph{Proc.
  of the 21st ACM MobiCom}, 2015, p. 65–76.

\bibitem{b4}
Z.~{Chen}, L.~{Zhang}, C.~{Jiang}, Z.~{Cao}, and W.~{Cui}, ``{WiFi CSI Based
  Passive Human Activity Recognition Using Attention Based BLSTM},'' \emph{IEEE
  Transactions on Mobile Computing}, vol.~18, no.~11, pp. 2714--2724, 2019.

\bibitem{b5}
W.~Jiang, C.~Miao, F.~Ma, S.~Yao, Y.~Wang, Y.~Yuan, H.~Xue, C.~Song, X.~Ma,
  D.~Koutsonikolas, W.~Xu, and L.~Su, ``{Towards Environment Independent Device
  Free Human Activity Recognition},'' in \emph{Proc. of the 24th ACM MobiCom},
  2018, p. 289–304.

\bibitem{b6}
Y.~{Zou}, W.~{Liu}, K.~{Wu}, and L.~M. {Ni}, ``{Wi-Fi Radar: Recognizing Human
  Behavior with Commodity Wi-Fi},'' \emph{IEEE Communications Magazine},
  vol.~55, no.~10, pp. 105--111, 2017.

\bibitem{b8}
F.~{Wang}, W.~{Gong}, and J.~{Liu}, ``{On Spatial Diversity in WiFi-Based Human
  Activity Recognition: A Deep Learning-Based Approach},'' \emph{IEEE Internet
  of Things Journal}, vol.~6, pp. 2035--2047, 2019.

\bibitem{b9}
H.~Huang and S.~Lin, ``{WiDet: Wi-Fi Based Device-Free Passive Person Detection
  with Deep Convolutional Neural Networks},'' in \emph{Proc. of the 21st ACM
  MSWIM}, 2018, p. 53–60.

\bibitem{b10}
X.~{Zheng}, J.~{Wang}, L.~{Shangguan}, Z.~{Zhou}, and Y.~{Liu}, ``{Smokey:
  Ubiquitous Smoking Detection with Commercial WiFi Infrastructures},'' in
  \emph{Proc. of IEEE INFOCOM}, 2016, pp. 1--9.

\bibitem{b11}
Y.~{Lin}, J.~{Le Kernec}, S.~{Yang}, F.~{Fioranelli}, O.~{Romain}, and
  Z.~{Zhao}, ``{Human Activity Classification With Radar: Optimization and
  Noise Robustness With Iterative Convolutional Neural Networks Followed With
  Random Forests},'' \emph{IEEE Sensors Journal}, vol.~18, no.~23, pp.
  9669--9681, 2018.

\bibitem{b7}
Y.~Wang, X.~Jiang, R.~Cao, and X.~Wang, ``{Robust Indoor Human Activity
  Recognition Using Wireless Signals},'' \emph{Sensors}, vol.~15, no.~7, p.
  17195–17208, 2015.

\bibitem{b13}
J.~{Bryan} and Y.~{Kim}, ``{Classification of Human Activities on UWB Radar
  using a Support Vector Machine},'' in \emph{Proc. of IEEE Antennas and
  Propagation Society International Symposium}, 2010, pp. 1--4.

\bibitem{b14}
M.~A. {Al Hafiz Khan}, R.~{Kukkapalli}, P.~{Waradpande}, S.~{Kulandaivel},
  N.~{Banerjee}, N.~{Roy}, and R.~{Robucci}, ``{RAM: Radar-based Activity
  Monitor},'' in \emph{Proc. of The 35th INFOCOM}, 2016, pp. 1--9.

\bibitem{b15}
K.~{He}, X.~{Zhang}, S.~{Ren}, and J.~{Sun}, ``{Deep Residual Learning for
  Image Recognition},'' in \emph{Proc. of IEEE/CVF CVPR}, 2016, pp. 770--778.

\bibitem{b16}
M.~{Zhao}, T.~{Li}, M.~A. {Alsheikh}, Y.~{Tian}, H.~{Zhao}, A.~{Torralba}, and
  D.~{Katabi}, ``{Through-Wall Human Pose Estimation Using Radio Signals},'' in
  \emph{Proc. of IEEE/CVF CVPR}, 2018, pp. 7356--7365.

\bibitem{b12}
{Youngwook Kim} and H.~{Ling}, ``{Human Activity Classification based on
  Micro-Doppler Signatures using an Artificial Neural Network},'' in
  \emph{Proc. of IEEE Antennas and Propagation Society International
  Symposium}, 2008, pp. 1--4.

\bibitem{enhancing}
T.~Zheng, Z.~Chen, S.~Ding, and J.~Luo, ``{Enhancing RF Sensing with Deep
  Learning: A Layered Approach},'' \emph{IEEE Communications Magazine}, 2021.

\bibitem{b50}
\BIBentryALTinterwordspacing
 [Online]. Available:
  \url{https://www.raspberrypi.org/products/raspberry-pi-zero/}
\BIBentrySTDinterwordspacing

\bibitem{b55}
N.~{Czink}, X.~{Yin}, H.~{OZcelik}, M.~{Herdin}, E.~{Bonek}, and B.~H.
  {Fleury}, ``{Cluster Characteristics in a MIMO Indoor Propagation
  Environment},'' \emph{IEEE Transactions on Wireless Communications}, vol.~6,
  pp. 1465--1475, 2007.

\bibitem{b43}
P.~{Wang}, P.~{Chen}, Y.~{Yuan}, D.~{Liu}, Z.~{Huang}, X.~{Hou}, and
  G.~{Cottrell}, ``{Understanding Convolution for Semantic Segmentation},'' in
  \emph{Proc. of IEEE WACV}, 2018, pp. 1451--1460.

\bibitem{b47}
J.~Zhang, Z.~Tang, M.~Li, D.~Fang, P.~Nurmi, and Z.~Wang, ``{CrossSense:
  Towards Cross-Site and Large-Scale WiFi Sensing},'' in \emph{Proc. of the
  24th MobiCom}, 2018, p. 305–320.

\bibitem{wirush}
\BIBentryALTinterwordspacing
 [Online]. Available: \url{http://www.wirush.ai/en}
\BIBentrySTDinterwordspacing

\bibitem{b56}
Y.~Xie, J.~Xiong, M.~Li, and K.~Jamieson, ``{MD-Track: Leveraging
  Multi-Dimensionality for Passive Indoor Wi-Fi Tracking},'' in \emph{Proc. of
  The 25th ACM MobiCom}, 2019.

\bibitem{b57}
``{IEEE Standard for Information Technology—Telecommunications and
  Information Exchange between Systems Local and Metropolitan Area
  Networks—Specific Requirements - Part 11: Wireless LAN Medium Access
  Control (MAC) and Physical Layer (PHY) Specifications},'' \emph{IEEE Std
  802.11-2016 (Revision of IEEE Std 802.11-2012)}, pp. 1--3534, 2016.

\bibitem{b31}
D.~Halperin, W.~Hu, A.~Sheth, and D.~Wetherall, ``{Tool Release: Gathering
  802.11n Traces with Channel State Information},'' \emph{SIGCOMM Comput.
  Commun. Rev.}, vol.~41, p.~53, 2011.

\bibitem{b51}
A.~Bhartia, B.~Chen, F.~Wang, D.~Pallas, R.~Musaloiu-E, T.~T.-T. Lai, and
  H.~Ma, ``{Measurement-Based, Practical Techniques to Improve 802.11ac
  Performance},'' in \emph{Proc. of the ACM IMC}, 2017, p. 205–219.

\bibitem{b40}
D.~Vasisht, S.~Kumar, and D.~Katabi, ``{Decimeter-Level Localization with a
  Single WiFi Access Point},'' in \emph{Proc. of 13th USENIX NSDI}, 2016, pp.
  165--178.

\bibitem{rockpi}
\BIBentryALTinterwordspacing
 [Online]. Available: \url{https://rockpi.org/}
\BIBentrySTDinterwordspacing

\bibitem{b17}
\BIBentryALTinterwordspacing
 [Online]. Available: \url{https://www.xethru.com/}
\BIBentrySTDinterwordspacing

\bibitem{b52}
F.~Adib, Z.~Kabelac, D.~Katabi, and R.~C. Miller, ``{3D Tracking via Body Radio
  Reflections},'' in \emph{Proc. of the 11th USENIX NSDI}, 2014, p. 317–329.

\bibitem{b19}
J.~Smith, \emph{Mathematics of the Discrete Fourier Transform (DFT): With Audio
  Applications}.\hskip 1em plus 0.5em minus 0.4em\relax BookSurge Publishing,
  2007.

\bibitem{b53}
C.~Shi, J.~Liu, H.~Liu, and Y.~Chen, ``{Smart User Authentication through
  Actuation of Daily Activities Leveraging WiFi-Enabled IoT},'' in \emph{Proc.
  of the 18th ACM Mobihoc}, 2017.

\bibitem{b41}
A.~G. Howard, M.~Zhu, B.~Chen, D.~Kalenichenko, W.~Wang, T.~Weyand,
  M.~Andreetto, and H.~Adam, ``{MobileNets: Efficient Convolutional Neural
  Networks for Mobile Vision Applications},'' \emph{arXiv preprint
  arXiv:1704.04861}, 2017.

\bibitem{b42}
A.~Krizhevsky, I.~Sutskever, and G.~E. Hinton, ``{ImageNet Classification with
  Deep Convolutional Neural Networks},'' \emph{Commun. ACM}, vol.~60, p.
  84–90, 2017.

\bibitem{b45}
\BIBentryALTinterwordspacing
 [Online]. Available: \url{https://www.tensorflow.org/}
\BIBentrySTDinterwordspacing

\bibitem{b54}
T.~Chen and C.~Guestrin, ``{XGBoost: A Scalable Tree Boosting System},'' in
  \emph{Proc. of the 22nd ACM SIGKDD}, 2016, p. 785–794.

\bibitem{b35}
\BIBentryALTinterwordspacing
``{FCC (GPO) Title 47, Section 15 of the Code of Federal Regulations SubPart F:
  Ultrawideband}.'' [Online]. Available: \url{https://www.ecfr.gov/}
\BIBentrySTDinterwordspacing

\bibitem{b36}
\BIBentryALTinterwordspacing
 [Online]. Available:
  \url{https://cdn.rohde-schwarz.com/pws/dl\_downloads/dl\_application/ \\
  application\_notes/1cm55/1CM55\_0e.pdf}
\BIBentrySTDinterwordspacing

\bibitem{b37}
\BIBentryALTinterwordspacing
 [Online]. Available:
  \url{https://www.xethru.com/blog/posts/xethru-radar-emission-comparison}
\BIBentrySTDinterwordspacing

\bibitem{b46}
M.~{Sandler}, A.~{Howard}, M.~{Zhu}, A.~{Zhmoginov}, and L.~{Chen},
  ``{MobileNetV2: Inverted Residuals and Linear Bottlenecks},'' in \emph{Proc.
  of IEEE/CVF CVPR}, 2018, pp. 4510--4520.

\bibitem{b21}
A.~M. {Khan}, Y.~. {Lee}, S.~Y. {Lee}, and T.~. {Kim}, ``{Human Activity
  Recognition via an Accelerometer-Enabled-Smartphone Using Kernel Discriminant
  Analysis},'' in \emph{Proc. of the 5th FutureTech}, 2010, pp. 1--6.

\bibitem{b22}
D.~Ilisei and D.~M. Suciu, ``{Human-Activity Recognition with Smartphone
  Sensors},'' in \emph{On the Move to Meaningful Internet Systems: OTM 2019
  Workshops}, 2020, pp. 179--188.

\bibitem{b23}
A.~{Anjum} and M.~U. {Ilyas}, ``{Activity Recognition using Smartphone
  Sensors},'' in \emph{Proc. of 10th IEEE CCNC}, 2013, pp. 914--919.

\bibitem{b26}
M.~K. {Fiaz} and B.~{Ijaz}, ``{Vision based Human Activity Tracking using
  Artificial Neural Networks},'' in \emph{Proc. of IEEE International
  Conference on Intelligent and Advanced Systems}, 2010, pp. 1--5.

\bibitem{b27}
Y.~{Shi}, Y.~{Tian}, Y.~{Wang}, and T.~{Huang}, ``{Sequential Deep Trajectory
  Descriptor for Action Recognition With Three-Stream CNN},'' \emph{IEEE
  Transactions on Multimedia}, vol.~19, pp. 1510--1520, 2017.

\bibitem{b58}
S.~Yun, Y.-C. Chen, H.~Zheng, L.~Qiu, and W.~Mao, ``{Strata: Fine-Grained
  Acoustic-Based Device-Free Tracking},'' in \emph{Proc. of the 15th MobiSys},
  2017, p. 15–28.

\bibitem{b59}
K.~Sun, T.~Zhao, W.~Wang, and L.~Xie, ``{VSkin: Sensing Touch Gestures on
  Surfaces of Mobile Devices Using Acoustic Signals},'' in \emph{Proc. of the
  24th ACM MobiCom}, 2018, p. 591–605.

\bibitem{b29}
A.~E. {Kosba}, A.~{Saeed}, and M.~{Youssef}, ``{Robust WLAN Device-free Passive
  Motion Detection},'' in \emph{Proc. of IEEE WCNC}, 2012, pp. 3284--3289.

\bibitem{b30}
H.~{Abdelnasser}, K.~{Harras}, and M.~{Youssef}, ``{A Ubiquitous WiFi-Based
  Fine-Grained Gesture Recognition System},'' \emph{IEEE Transactions on Mobile
  Computing}, vol.~18, pp. 2474--2487, 2019.

\bibitem{ziqi_fall_2018}
Z.~Wang, Z.~Gu, J.~Yin, Z.~Chen, and Y.~Xu, ``{Syncope Detection in Toilet
  Environments Using Wi-Fi Channel State Information},'' in \emph{Proc. of the
  ACM UbiComp}, 2018, p. 287–290.

\bibitem{b32}
S.~Sen, J.~Lee, K.-H. Kim, and P.~Congdon, ``{Avoiding Multipath to Revive
  Inbuilding WiFi Localization},'' in \emph{Proc. of the 11th ACM MobiSys},
  2013, p. 249–262.

\bibitem{zhe_rfnet_2020}
S.~Ding, Z.~Chen, T.~Zheng, and J.~Luo, ``{RF-Net: A Unified Meta-Learning
  Framework for RF-Enabled One-Shot Human Activity Recognition},'' in
  \emph{Proc. of the 18th ACM SenSys}, 2020, p. 517–530.

\bibitem{b33}
M.~Zhao, Y.~Tian, H.~Zhao, M.~A. Alsheikh, T.~Li, R.~Hristov, Z.~Kabelac,
  D.~Katabi, and A.~Torralba, ``{RF-Based 3D Skeletons},'' in \emph{Proc. of
  the ACM SIGCOMM}, 2018, p. 267–281.

\bibitem{m3}
Z.~Chen, G.~Zhu, S.~Wang, Y.~Xu, J.~Xiong, J.~Zhao, J.~Luo, and X.~Wang,
  ``{M$^3$: Multipath Assisted Wi-Fi Localization with a Single Access
  Point},'' \emph{IEEE Transactions on Mobile Computing}, vol.~20, no.~2, pp.
  588--602, 2020.

\bibitem{luo2014ilocscan}
C.~Zhang, F.~Li, J.~Luo, and Y.~He, ``{iLocScan: Harnessing Multipath for
  Simultaneous Indoor Source Localization and Space Scanning},'' in \emph{Proc.
  of the 12th ACM SenSys}, 2014, p. 91–104.

\bibitem{zheng2020v2ifi}
T.~Zheng, Z.~Chen, C.~Cai, J.~Luo, and X.~Zhang, ``{V$^2$iFi: in-Vehicle Vital
  Sign Monitoring via Compact RF Sensing},'' in \emph{Proc. of the 20th ACM
  UbiComp}, 2020, pp. 1--27.

\bibitem{UWHear}
Z.~Wang, Z.~Chen, A.~Singh, L.~Garcia, J.~Luo, and M.~Srivastava, ``{UWHear:
  Through-wall Extraction and Separation of Audio Vibrations Using Wireless
  Signals},'' in \emph{Proc. of the 18th ACM SenSys}, 2020, pp. 1--14.

\end{thebibliography}

\begin{IEEEbiography}[{\includegraphics[width=1in,height=1.2in,clip,keepaspectratio]{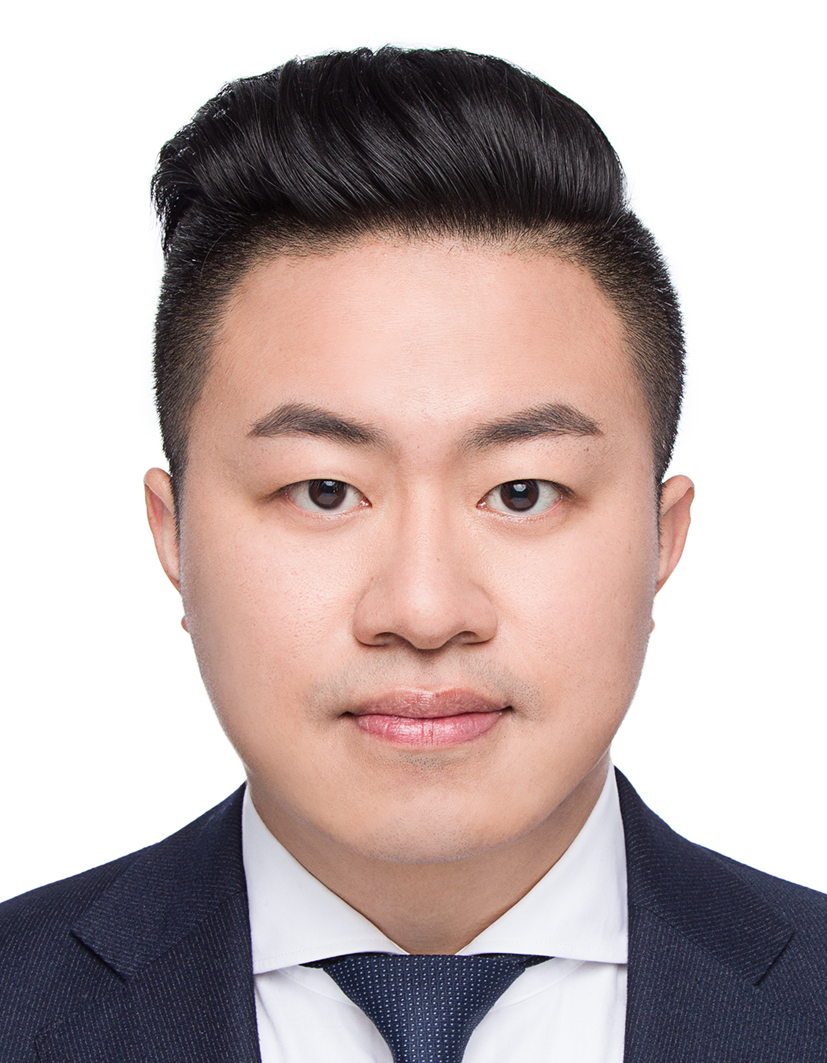}}]{Zhe Chen}
(chen.zhe@ntu.edu.sg) is a research fellow in Nanyang Technological University, Singapore. He received the Ph.D. degree with honor in Computer Science from Fudan University, Shanghai, China, and obtained Doctoral Dissertation Award from ACM SIGCOMM China 2019. His research interests include RF communication and sensing systems, deep learning, and IoT applications.
\end{IEEEbiography}

\begin{IEEEbiography}[{\includegraphics[width=1in,height=1.2in,clip,keepaspectratio]{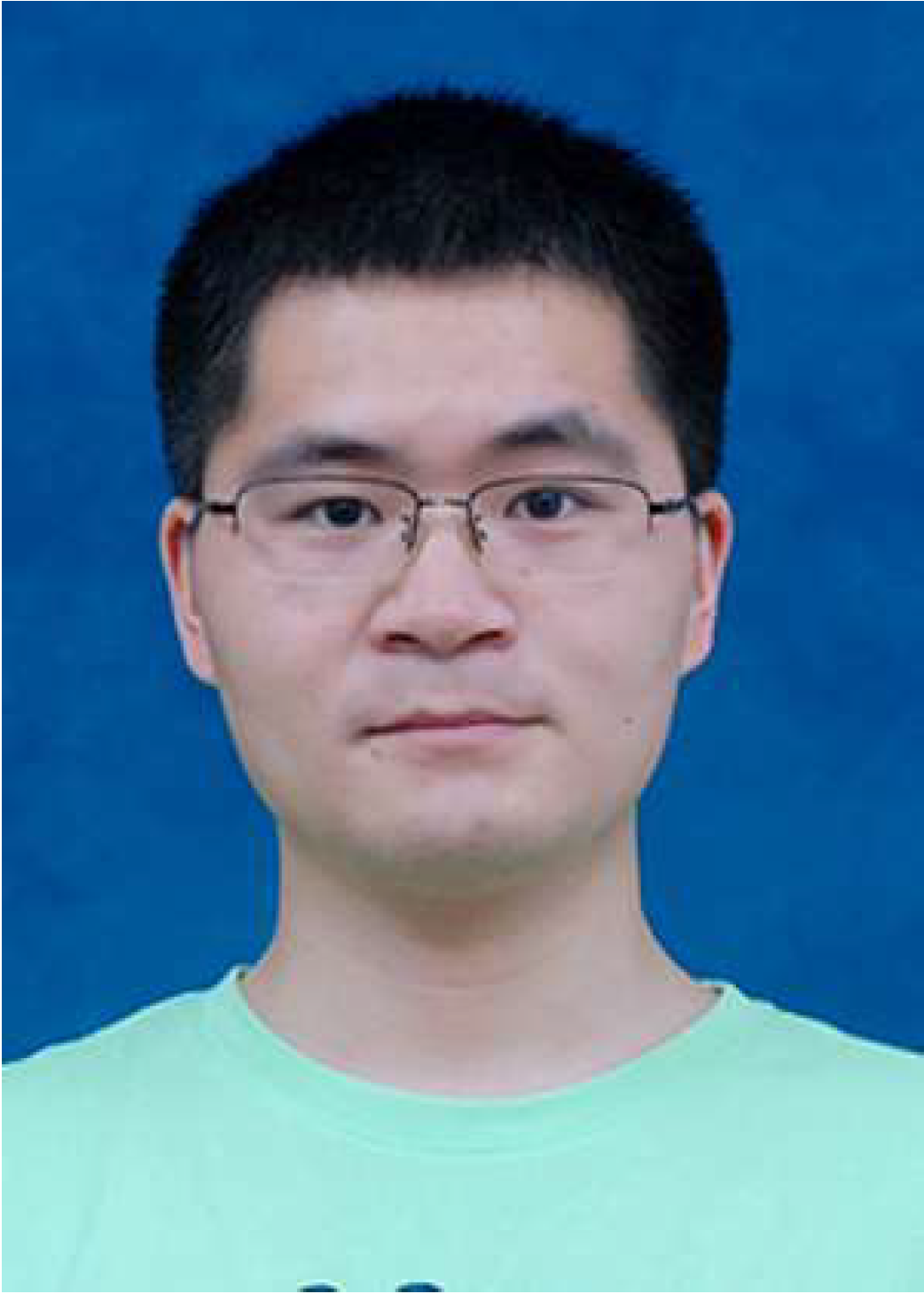}}]{Chao Cai} is a research fellow with the School of Computer Science and Engineering at Nanyang Technological University, Singapore. He got his MS and Ph.D. at School of Electronic Information and Engineering, Huazhong University of Science and Technology. 
His current research interests include mobile computing, acoustic sensing, wireless sensing, embedded system, digital signal processing, and deep learning.
\end{IEEEbiography}

\begin{IEEEbiography}[{\includegraphics[width=1in,height=1.2in,clip,keepaspectratio]{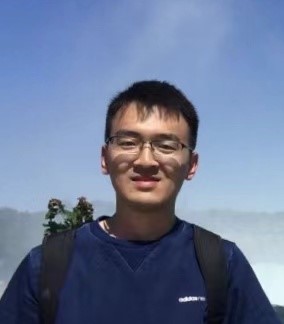}}]{Tianyue Zheng}(tianyue002@e.ntu.edu.sg) received his B.Eng. degree in Telecommunication Engineering from Harbin Institute of Technology, China, and M.Eng. degree in Computer Engineering from the University of Toronto, Canada. He is currently working towards his Ph.D. degree in Computer Science from Nanyang Technological University, Singapore. His research interests include RF sensing and deep learning.
\end{IEEEbiography}

\begin{IEEEbiography}[{\includegraphics[width=1in,height=1.2in,clip,keepaspectratio]{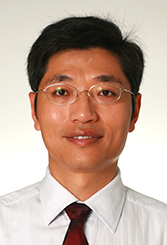}}]{Jun Luo} received his BS and MS degrees in
Electrical Engineering from Tsinghua University,
China, and the Ph.D. degree in Computer Science from EPFL (Swiss Federal Institute of Technology in Lausanne), Lausanne, Switzerland.
From 2006 to 2008, he has worked as a postdoctoral research fellow in the Department of
Electrical and Computer Engineering, University of Waterloo, Waterloo, Canada. In 2008, he
joined the faculty of the School Of Computer Science and Engineering, Nanyang Technological
University in Singapore, where he is currently an Associate Professor.
His research interests include mobile and pervasive computing, wireless
networking, machine learning and computer vision, as well as applied operations research.
More information can be found at http://www.ntu.edu.sg/home/junluo.
\end{IEEEbiography}

\begin{IEEEbiography}[{\includegraphics[width=1in,height=1.2in,clip,keepaspectratio]{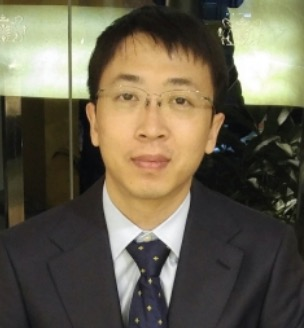}}]{Jie Xiong}
received the B.Eng. degree from Nanyang
Technological University, Singapore, in 2005, the
M.Sc. degree from Duke University, Durham, NC,
USA, in 2009, and the Ph.D. degree in computer science from University College London, London, U.K.,
in 2015.
He is an Assistant Professor in the College of Information and Computer Sciences
UMass Amherst, USA. His research interests include building practical wireless and mobile systems that bridge
the gaps between theory and reality. His recent work
appears at MobiCom, NSDI, CoNEXT, Ubicomp, and INFOCOM.
Prof. Xiong was the recipient of the prestigious Google European Doctoral
Fellowship in Wireless Networking for his doctoral studies. His Ph.D. thesis was the 2016 British Computer Society Distinguished Dissertation Award
runner-up.
\end{IEEEbiography}

\begin{IEEEbiography}[{\includegraphics[width=1in,height=1.2in,clip,keepaspectratio]{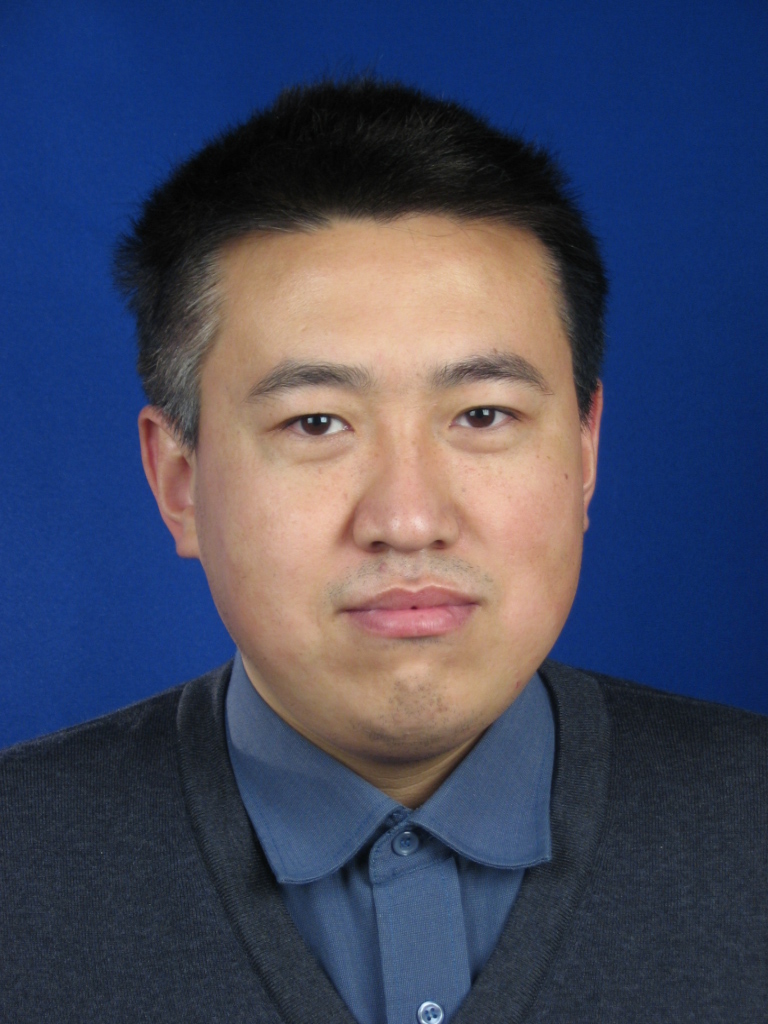}}]{Xin Wang}
is a professor at Fudan University,
Shanghai, China. He received his BS Degree
in Information Theory and MS Degree in Communication
and Electronic Systems from Xidian
University, China, in 1994 and 1997, respectively.
He received his Ph.D. Degree in Computer
Science from Shizuoka University, Japan,
in 2002. His research interests include quality
of network service, next-generation network architecture,
mobile Internet and network coding.
Contact him at xinw@fudan.edu.cn.
\end{IEEEbiography}

\vfill

\end{document}